\documentclass[11pt]{article} 

\makeatletter\renewcommand{\section}{\@startsection
	{section}{1}{\z@}{-3.5ex plus -1ex minus
		-.2ex}{2.3ex plus .2ex}{\bf }}

\makeatletter\renewcommand{\subsection}{\@startsection{subsection}{2}{\z@}{-3.25ex
		plus -1ex minus
		-.2ex}{1.5ex plus .2ex}{\it }}
\makeatletter\renewcommand{\subsubsection}{\@startsection{subsubsection}{3}{-2.45ex}{-3.25ex
		plus -1ex minus -.2ex}{1.5ex plus .2ex}{\it }}
\renewcommand{\thesection}{\arabic{section}}
\renewcommand{\thesubsection}{\arabic{section}.\arabic{subsection}.}

\renewcommand{\theequation}{\thesection.\arabic{equation}}
\makeatletter \@addtoreset{equation}{section}

\usepackage{graphicx}
\graphicspath{{plots/}} 
\usepackage{float} 
\usepackage{epsfig} 
\usepackage{verbatim}
\usepackage{hyperref}
\usepackage{easyfig}
\usepackage{bbm}
\usepackage{amsmath}
\usepackage{amsfonts}
\usepackage[compact]{titlesec}
\usepackage{subdepth}
\usepackage{dsfont}
\usepackage{braket}
\usepackage{tikz}
\usepackage[utf8]{inputenc}
\usepackage[T1]{fontenc}
\usepackage{amsmath}
\usepackage[toc]{glossaries}

\usepackage{caption}
\usepackage{subcaption}

\renewenvironment{thebibliography}[1]
{\baselineskip=16pt plus 2pt minus 1pt
	\section*{\large\refname
		\@mkboth{\MakeUppercase\refname}{\MakeUppercase\refname}}%
	\list{\@biblabel{\@arabic\c@enumiv}}%
	{\settowidth\labelwidth{\@biblabel{#1}}%
		\leftmargin\labelwidth
		\advance\leftmargin\labelsep
		\@openbib@code
		\usecounter{enumiv}%
		\let\p@enumiv\@empty
		\renewcommand\theenumiv{\@arabic\c@enumiv}}%
	\sloppy
	\clubpenalty4000
	\@clubpenalty \clubpenalty
	\widowpenalty4000%
	\sfcode`\.\@m}

\let\fn\footnote
\renewcommand{\footnote}[1]{\linespread{1.1}\fn{#1}\linespread{1.29}}

\usepackage{physics}
\usepackage{amsfonts}
\usepackage{amssymb}
\usepackage{mathrsfs}
\usepackage{amsmath,amssymb}
\usepackage{bm}
\usepackage{caption}
\usepackage{amsmath}
\usepackage{graphicx}
\usepackage{cite}

\usepackage[T1]{fontenc}
\usepackage[utf8]{inputenc}

\newcommand{\appendices}{\section*{Appendices}\setcounter{section}{0} \setcounter{equation}{0}
	\renewcommand{\thesection}{\Alph{section}.}
	\renewcommand{\thesubsection}{\Alph{section}.\arabic{subsection}.}
	\renewcommand{\theequation}{\thesection\arabic{equation}}}

\def\tyng(#1){\hbox{\tiny$\yng(#1)$}}

\newcommand{\be}{\begin{equation}}
	\newcommand{\ee}{\end{equation}}
\newcommand{\bea}{\begin{array}}
	\newcommand{\ea}{\end{array}}
\newcommand{\beqa}{\begin{eqnarray}}
	\newcommand{\eeqa}{\end{eqnarray}}
\newcommand{\nn}{\nonumber}

\usepackage{pdfpages}

\begin{document}
	
	\fontfamily{bch}\fontsize{11pt}{15pt}\selectfont
	\begin{titlepage}
		\begin{flushright}
		\end{flushright}
		
		
		\begin{center}
			{\Large \bf Chaos in $SU(2)$ Yang-Mills Chern-Simons Matrix Model }\\
			~\\
			
			
			\vskip 3em
			
			\centerline{$ \text{\large{\bf{K. Ba\c{s}kan}}} \, \, $, $ \text{\large{\bf{S. K\"{u}rk\c{c}\"{u}o\v{g}lu}}}$}
			\vskip 0.5cm
			\centerline{\sl Middle East Technical University, Department of Physics,}
			\centerline{\sl Dumlupınar Boulevard, 06800, Ankara, Turkey}
			
			\vskip 1em
			
			\begin{tabular}{r l}
				E-mails: 
				&\!\!\!{\fontfamily{cmtt}\fontsize{11pt}{15pt}\selectfont kagan.baskan@metu.edu.tr  }\\
				&\!\!\!{\fontfamily{cmtt}\fontsize{11pt}{15pt}\selectfont kseckin@metu.edu.tr}
				
			\end{tabular}
			
		\end{center}
		
		\vskip 5 em
		
		\begin{quote}
			\begin{center}
				{\bf Abstract}
			\end{center}
				
				We study the effects of addition of Chern-Simons (CS) term in the minimal Yang Mills (YM) matrix model composed of two $2 \times 2$ matrices with $SU(2)$ gauge and $SO(2)$ global symmetry. We obtain the Hamiltonian of this system in appropriate coordinates and demonstrate that its dynamics is sensitive to the values of both the CS coupling, $\kappa$, and the conserved conjugate momentum, $p_\phi$, associated to the $SO(2)$ symmetry. We examine the behavior of the emerging chaotic dynamics by computing the Lyapunov exponents and plotting the Poincar\'{e} sections as these two parameters are varied and, in particular, find that the largest Lyapunov exponents evaluated within a range of values of $\kappa$ are above that is computed at $\kappa=0$, for $\kappa p_\phi < 0$. We also give estimates of the critical exponents for the Lyapunov exponent as the system transits from the chatoic to non-chaotic phase with $p_\phi$ approaching to a critical value.

		\vskip 1em

	\end{quote}
		
	\end{titlepage}
	
	\setcounter{footnote}{0}
	\pagestyle{plain} \setcounter{page}{2}
	
	\newpage	
	
\section{Introduction}

Recently, there has been growing interest in exploring the structure of chaotic dynamics emerging from the matrix quantum mechanics \cite{Sekino:2008he,  Asplund:2011qj,  Shenker:2013pqa, Gur-Ari:2015rcq, Berenstein:2016zgj, Maldacena:2015waa, Aoki:2015uha, Asano:2015eha, Berkowitz:2016jlq, Buividovich:2017kfk, Buividovich:2018scl, Coskun:2018wmz}, such as the BFSS and the BMN models \cite{Banks:1996vh, deWit:1988wri, Itzhaki:1998dd,  Berenstein:2002jq, Dasgupta:2002hx, Ydri:2017ncg,Ydri:2016dmy} which appear in the DLCQ quantization of M-theory in the flat and the pp-wave background, respectively. These models are $SU(N)$ gauge theories, describing the dynamics of the $N$-coincident $D0$-branes, in the flat and spherical backgrounds. It is well-known that the gravity dual is obtained in the 't Hooft limit, i.e. at large $N$ and strong Yang-Mills (YM) coupling and describes a phase in which $D0$-branes form a so called black-brane, i.e. a string theoretical black hole \cite{Ydri:2017ncg,Ydri:2016dmy,Kiritsis:2007zza}. While the earlier investigations (and some recent as well  \cite{Kawahara:2006hs, Kawahara:2007fn, DelgadilloBlando:2007vx, DelgadilloBlando:2008vi, Asano:2018nol, Berkowitz:2018qhn} ) on the quantum mechanical behaviour of these models were performed in the Euclidean time formulation using both analytical perturbative and Monte-Carlo methods, in the past few years, there has been increasing interest on accessing the quantum dynamics using real-time formulations \cite{Buividovich:2017kfk, Buividovich:2018scl}. These studies are propelled by a result due Maldacena-Shenker-Stanford (MSS)  \cite{Maldacena:2015waa}, which states that under general circumstances, the Lyapunov exponent (which is a measure of chaos in both classical and quantum mechanical systems) for quantum chaos is bounded, that this bound is controlled by the temperature of the system, and given by $\lambda_L \leq 2\pi T$. It is conjectured that systems which are holographically dual  to the black holes, are expected to be maximally chaotic. This is already demonstrated for the Sachdev-Ye-Kitaev (SYK) \cite{Maldacena:2016hyu}  model, and expected to be so for the BFSS model too. Numerical studies reported in \cite{Gur-Ari:2015rcq} found that, for the BFSS model treated at the classical level, the largest Lyapunov exponent is given as $\lambda_L = 0.2924(3) (\lambda_{'t \, Hooft})^{1/4}$. This is parametrically smaller than the MSS bound $2\pi T$ and violates it only temperatures below $\approx 0.015$, while the quantum correction recently evaluated using Gaussian state approximation \cite{Buividovich:2018scl}, indicate that the largest Lyapunov exponent vanishes below a non-zero temperature, and hence ensuring that the MSS bound is not violated.

It is important to note that not only the BFSS, BMN matrix models, but even their subsectors at small values of $N$ appear as non-trivial many-body systems, and we lack a complete solution to these or even for the smallest Yang-Mills (YM) matrix model to date. The latter may be described as being composed of two $2\times 2$ Hermitian matrices with $SU(2)$ gauge and $SO(2)$ global symmetries. It can be obtained by dimensionally reducing the YM theory from $2+1$- to $0+1$-dimensions. Classical dynamics of this system was recently investigated in \cite{Berenstein:2016zgj} (see also the references \cite{Kabat:1996cu, Kares:2004uk} in this context) and it was shown that, using the $SU(2)$ gauge and $SO(2)$ rotations of the two matrices among themselves and a judicious choice of coordinates to fully implement the Gauss law constraint leads to a Hamiltonian with two degrees of freedom and their conjugate momenta. In addition, the angular momentum, $p_\phi$, associated to the rigid $SO(2)$ symmetry appears as a conserved quantity via a term proportional to the square of $p_\phi$ and strongly controls the structure of the effective potential and the ensuing dynamics. At $p_\phi = 0$, the model collapses to the usual $x^2y^2$ potential, which is already known to lead to almost completely chaotic dynamics \cite{Matinyan:1981dj, Savvidy:1982wx, Savvidy:1982jk, Arefeva:1997oyf}. In \cite{Berenstein:2016zgj}, the response of the system to a range of different values of $p_\phi$ is investigated and it is found that, at fixed energy, there is a value of $p_\phi$ above which the chaos ceases to exist and the dynamics is essentially described by quasi-periodic motion. Therefore, the model is conjectured to have two phases, namely a chaotic phase corresponding to a toy model for a black hole, and a phase consisting of two $D0$-branes tied with fixed number of open strings stretching between them, with a force that depends on the number of excited strings. The latter can be roughly thought of as the "adiabatic invariant" for the quasi-periodic orbits, which appear as the Kolmogorov-Arnold-Moser (KAM) tori (see, for instance, \cite{Ott}) in the Poincar\'e sections. For a given value of energy these two phases can coexist within a range of values of $p_\phi$, while the end of chaotic dynamics is argued to correspond to the end of the black hole phase. Quantum aspects of the $2 \times 2$ matrix model is addressed in \cite{Hubener:2014pfa}, where the ground state energy is also estimated. 

In order to gain further insight into the matrix model composed of $2\times 2$ matrices with $SU(2)$ gauge symmetry, in this paper, we set out to investigate the dynamics in the presence of the Chern-Simons (CS) term. It is possible to obtain the corresponding action starting from the $SU(2)$ Yang-Mills Chern-Simons (YMCS) model in $2+1$ dimensions and reducing it to $0+1$\footnote{Let us immediately note here that, although the CS coupling is quantized for the non-abelian CS term in $2+1$ dimensions, this is not so after dimensional reduction to $0+1$ since the coupling of this model involves the $2$-dimensional volume factor, and CS term is indeed gauge invariant in $0+1$ dimensions. Full details of this reduction and related facts are provided in the appendix A.}. In a manner similar to the one followed in \cite{Berenstein:2016zgj}, while paying attention to the differences in the procedure due to the CS term, which is first order in time derivative, we obtain the Hamiltonian of the system. The latter has the same degrees of freedom as the pure YM model, while the effective potential is governed not only by $p_\phi$, but also the CS coupling $\kappa$, which enters into the effective potential via $\kappa p_\phi$ and another term $\propto \kappa^2 r^2$. Varying $\kappa$ at different values of $p_\phi$, we probe the impact on the chaotic dynamics. Our new findings are as follows. Firstly, we find that at $p_\phi=0$, values of the largest (and only) Lyapunov exponent are above that evaluated at $\kappa=0$, approximately within the range of values of $4 \pi |\kappa| \lesssim 4$. This can be attributed to shrinking of the sharp edges of the effective potential contours (see figure \ref{fig:ctrpf0fig1}), but not sustained further for $4 \pi |\kappa|\gtrsim 4$ as the harmonic term $\propto \kappa^2 r^2$ starts to dominate and chaotic dynamics gradually declines. The second and more interesting effect is due to the $\kappa p_\phi$ term, which alters the Lyapunov spectrum depending on its sign, in other words, the orientation of $p_\phi$ matters. For instance, we find the values of the largest Lyapunov exponent for $\kappa p_\phi < 0$  for a range of values of $\kappa $ at fixed $p_\phi$ are above that is evaluated at $\kappa=0$. These results are presented and discussed in detail in section 3, where our findings obtained from the Lyapunov data are further corroborated via the use of Poincar\'{e} sections.  As another important finding, we give estimates for the critical exponents for $\lambda_L$ and the value of the order parameter, $p_\phi^c$, as the system transits from chaotic to non-chaotic phase. Rest of the paper is organized as follows. Section 2 gives the developments leading to the Hamiltonian of the model. Most of the details of the calculations in this section are relegated to the appendices for completeness. We summarize our results and briefly state our conclusions in section 4.

\section{$SU(2)$ Matrix Model with the Chern-Simons Term}

The action of the model may be given as
\be
S = S_{YM} + S_{CS} \,,
\label{actionYMCS}
\ee
where
\begin{equation} 
\label{actionYM}
S_{YM} = \int{\dd t \Tr[\dfrac{1}{2}(D_0X_i)^2+\dfrac{1}{4}\comm{X_i}{X_j}^2]} \,, 
\end{equation}
and
\begin{equation}
S_{CS} = \kappa \int dt \, \Tr \lbrack \epsilon_{ij} (X_i \dot{X}_j + 2i A_0 X_i X_j) \rbrack = \kappa \int dt \, \Tr \lbrack \epsilon_{ij} X_i  \,( D_0 X_j) \rbrack \,.
\label{actionCS}
\end{equation}
In these expressions $X_1, X_2$ are $2 \times 2$ traceless Hermitian matrices whose entries are functions of time only. They transform under the adjoint representation of $SU(2)$: $X_i \rightarrow U^\dagger X_i U$ as usual. $D_0 X_i = \partial_0 X_i - i \lbrack A_0 , X_i \rbrack $ are the covariant derivatives, and $A_0$ is a gauge field which transforms accordingly under the local $SU(2)$ gauge group. $S$ is invariant under the local $SU(2)$ gauge symmetry as well as under a global $SO(2)$; i.e. the "rigid" rotations of the $X_i$'s among themselves. In (\ref{actionCS}), $\kappa$ is the CS coupling constant. Note that,  due the gauge invariance of $S_{CS}$ term in $0+1$-dimensions, $\kappa$ is not level quantized\footnote{A detailed discussion is provided in the appendix A.}. Let us also note that we have implicitly set the Yang-Mills coupling, an overall factor $\frac{1}{g^2}$ in $S_{YM}$, to unity. YM coupling can easily be restored back in the action by performing the scalings $t \rightarrow g^{\frac{2}{3}} t $, $\partial_0 \rightarrow g^{-\frac{2}{3}} \partial_0 $, $A_0 \rightarrow g^{-\frac{2}{3}} A_0 $, $X_i \rightarrow g^{-\frac{2}{3}} X_i $ and $\kappa \rightarrow g^{\frac{4}{3}} \kappa$. Pure CS model limit is obtained by letting $g \rightarrow \infty$ and it is discussed in detail in the appendices C and D.

It is convenient to work in the $A_0 = 0$ gauge. In the presence of the CS term, Gauss law constraint takes the form 
\begin{equation}
-\comm{X_i}{\dot{X}_i} + 2\kappa\epsilon_{ij}X_iX_j = 0 \,.
\label{gausslaw}
\end{equation}

We may express the matrices $X_i$ as
\begin{equation}
X_i = \frac{1}{\sqrt{2}} \va{x}_i \vdot \va{\sigma} = \frac{1}{\sqrt{2}} x_i^\alpha \sigma^\alpha, \quad i:1,2, \quad \& \quad  \alpha=1,2,3,
\label{matexp1}
\end{equation}
where $\frac{1}{\sqrt{2}}$ is a normalization factor and $\sigma^\alpha$ are the usual Pauli matrices. For future notational convenience, it is also useful to arrange components of $X_i$ into column vectors
\begin{align}
\begin{split}
\va{x}_1 = 
\begin{pmatrix}
x_1^1 \\
x_1^2 \\
x_1^3
\end{pmatrix} \,, \quad 
\va{x}_2 =
\begin{pmatrix}
x_2^1 \\
x_2^2 \\
x_2^3 
\end{pmatrix} \,.
\end{split}
\label{cvectors}
\end{align}

Substituting (\ref{matexp1}) into the action (\ref{actionYMCS}) yields the Lagrangian 
\begin{equation}
L = \frac{1}{2}(\dot{\va{x}}_1^2 + \dot{\va{x}}_2^2)  + \kappa(\va{x}_1 \vdot \dot{\va{x}}_2 - \va{x}_2 \vdot \dot{\va{x}}_1) - (\va{x}_1 \cross \va{x}_2)^2 \,,
\label{carLag}
\end{equation}
while the constraint (\ref{gausslaw}) takes the form
\begin{equation}
\va{x}_1 \cross \dot{\va{x}}_1 + \va{x}_2\cross\dot{\va{x}}_2 - 2\kappa \, \va{x}_1\cross\va{x}_2 = 0 \,.
\label{gausslaw2}
\end{equation}
The canonical conjugate momenta are easily obtained from the Lagrangian (\ref{carLag}) as
\begin{align}\
\begin{split}
\va{p}_1 = \dot{\va{x}}_1 - \kappa\va{x}_2 \,, \\
\va{p}_2 = \dot{\va{x}}_2 + \kappa\va{x}_1 \,,
\end{split}
\label{cmomenta}
\end{align}
which clearly show that the kinematical and conjugate momenta are no longer the same in the presence of the CS term; a fact which is widely known in the literature (see for instance, \cite{Dunne}). Defining
\be
\va{L}_1 = \va{x}_1  \times \va{p}_1 \,,\quad \va{L}_2 = \va{x}_2  \times \va{p}_2 \,,
\ee
Gauss law constraint in (\ref{gausslaw2}) can be expressed as the condition of the vanishing of the  $SU(2)$ angular momentum:
\begin{equation}
\va{L} := \va{L}_1 + \va{L}_2 = 0 \,.
\label{gausslaw3}
\end{equation}

In order to obtain the corresponding Hamiltonian, we need to observe that the Lagrangian involves a term which is first order in time derivatives. Let us note that the generic form of such a Lagrangian can be given as 
\begin{equation}
L(q_a, \dot{q}_a, t) = \frac{1}{2} g_{ab} \dot{q}_a \dot{q}_b + f_a \dot{q}_a - V \,, \quad a,b:1\,,\cdots\,, K \,,
\label{genericLag}
\end{equation}
where $g_{ab}$ is the metric associated to the generalized coordinates $q_a$,  $f_a$ are functions of the generalized coordinates, i.e. $f_a\equiv f_a(q_b)$ and $V \equiv V(q_a)$ is the potential. Corresponding Hamiltonian can be shown to take the form (see the appendix C for details)
\begin{equation}
H = \frac{1}{2}g_{ab}^{-1}p_a p_b + \frac{1}{2}g_{ab}^{-1} f_a f_b - g_{ab}^{-1} f_a p_b + V \,.
\label{genericham}
\end{equation}
Adapting (\ref{genericham}) to (\ref{carLag}), in the Cartesian coordinates we obviously have $g_{ab}$ as the Euclidean flat metric $\delta_{ij}$, we may write  $\va{f}_i = - \kappa \varepsilon_{ij} \va{x}_j$ ($i,j:1,2$) and observe that $V = (\va{x}_1 \cross \va{x}_2)^2$. Putting all these together, we find that the Hamiltonian corresponding to (\ref{carLag}) takes the form\footnote{In the pure CS limit the Hamiltonian becomes zero as explained in the appendices C and D.}
\begin{equation}
H=\frac{1}{2}(\va{p}_1^2 + \va{p}_2^2) + \frac{1}{2}\kappa^2(\va{x}_1^2 + \va{x}_2^2) + \kappa(\va{p}_1\vdot\va{x}_2 - \va{p}_2\vdot\va{x}_1) + (\va{x}_1 \cross \va{x}_2)^2 \,,
\label{Hamiltonian}
\end{equation}
with the  equations of motion easily evaluated to be
\begin{align}
\begin{split}
\dot{\va{x}}_1 &= p_1 \,, \\
\dot{\va{p}}_1 &= - \kappa^2 \va{x}_1 + \kappa\va{p}_2 - 2 \va{x}_2\cross(\va{x}_1\cross\va{x}_2) \,, \\
\dot{\va{x}}_2 &=  p_2 \,, \\
\dot{\va{p}}_2 &= - \kappa^2 \va{x}_2 - \kappa\va{p}_1 + 2 \va{x}_2\cross(\va{x}_1\cross\va{x}_2) \,.
\end{split}
\label{Heom}
\end{align}
  
Using (\ref{Heom}), the time derivative of $\va{L}_1$ may be expressed as 
\begin{equation}
\dot{\va{L}}_1 = \kappa \, \va{x}_1\cross\va{p}_2 - 2\va{x}_1\cross(\va{x}_2\cross(\va{x}_1\cross\va{x}_2)) \,.
\label{l1dot}
\end{equation}
A similar result holds for $\dot{\va{L}}_2$. Although the second term in (\ref{l1dot}) remains aligned with $\va{L}_1$ as it does in the pure YM matrix model, this is not manifest for the first term. Nevertheless, the subsequent analysis will show, upon implementing the Gauss law in appropriate coordinates, that the dynamics remain planar.

Taking advantage of the local $SU(2) \approx SO(3) $ and the global $SO(2)$ rotations, we may introduce the coordinates $(\alpha, \beta, \gamma, r, \theta, \phi)$. Following \cite{Berenstein:2016zgj}, we may consider the $3 \times 2$ matrix $M$ whose columns are the vectors $\va{x}_1$ and $\va{x}_2$ , i.e. $M = (\va{x}_1, \va{x}_2)$ and express $M$ as
\begin{equation}
M = \frac{1}{\sqrt{2}} \, R(\alpha,\beta,\gamma)
\vdot 
\begin{pmatrix}
r &  r \cos \theta  \\
0 & r \sin \theta \\
0 & 0
\end{pmatrix}
\vdot
\begin{pmatrix}
\cos \phi & \sin \phi \\
-\sin \phi & \cos \phi
\end{pmatrix} \,,
\label{newcoord}
\end{equation}
where $R(\alpha,\beta,\gamma)$ is a $SO(3)$ Euler matrix using  $z-x-z$ active rotation with the angles $(\alpha, \beta, \gamma)$, respectively. Its explicit form is given in the appendix for quick reference. $M^0 \equiv (\va{x}_1^{T,(0)}, \va{x}_2^{T,(0)})$ with $\va{x}_1^{T,(0)} := (r,0,0)$ and $\va{x}_2^{T,(0)} := (r \cos \theta, r \sin \theta, 0 ) $ may be thought as a configuration  of the two $D0$-branes oriented coplanarly with a relative angle $\theta$ obtained via a $SU(2)$ gauge choice. The latter is not preserved in general by the global $SO(2)$ rotations on $\va{x}_1$ and $\va{x}_2$, which can be taken to act on the right of $M^0$, nor it is preserved by the $SU(2) \approx SO(3)$ gauge rotations, which acts from the left on $M^0$. Thus, taking these facts together, (\ref{newcoord}) is a convenient way to introduce new coordinates for the present dynamical system. The advantage of this choice of the coordinates is that, the Gauss law constraint  in (\ref{gausslaw3}) can be fully solved and manifestly imposed on the Hamiltonian expressed in terms of the new variables, as we will demonstrate in what follows. Let us also remark that, this is essentially the same approach followed in \cite{Berenstein:2016zgj} except that, we no longer restrict the gauge $SU(2) \approx SO(3) $ rotations to an $SO(2)$ subgroup in advance, since, it is not readily seen that $\dot{\va{L}}_i$ $(i=1,2)$ remain aligned with  $\va{L}_i$.
 
Let us note in advance that the components of angular momentum $\va{L}$ can be expressed in terms of the conjugate momenta $(p_\alpha, p_\beta, p_\gamma)$ corresponding to the Euler angles $(\alpha, \beta, \gamma)$\footnote{As this is not frequently encountered in the literature, we provide a quick derivation in the appendix E.} as \cite{Marsden}
\begin{equation}
\va{L}=
\begin{pmatrix}
\sin \alpha (p_\gamma \csc \beta  - p_\alpha \cot \beta )+p_\beta \cos \alpha \\
\cos \alpha \csc \beta (p_\alpha \cos \beta - p_\gamma)+p_\beta \sin \alpha \\
p_\alpha
\label{angmomeuler}
\end{pmatrix}\,.
\end{equation}
which immediately implies that the Gauss law constraint  $\va{L} = 0$ is equivalent to
\be
p_\alpha = p_\beta = p_\gamma =0 \,.
\label{gausslaw4}
\ee
We will make use of (\ref{gausslaw4}) to fully impose the Gauss law in what follows.

The metric in the new coordinates $(r, \theta, \phi, \alpha, \beta, \gamma)$ is straightforwardly obtained from the expression
\begin{equation}
g_{ij} = \Tr(\partial_i M^\dagger \partial_j M) \,.
\label{inducedmetric}
\end{equation}
We give the  components of $g_{ij}$ and its inverse $g^{ij}$ in appendix D and also provide there the details of the evaluation of the Hamiltonian in the new coordinates
using the generic form in (\ref{genericham}) together with the inverse metric $g^{ij}$. Employing these facts and imposing the Gauss law constraint \eqref{gausslaw4}, we find 
\beqa
H &=& \frac{1}{2}p_r^2 + \frac{2}{r^2}p_\theta^2 + \frac{p_\phi^2}{2r^2\cos[2](\theta)} + \kappa p_\phi  + \frac{\kappa^2 r^2}{2} + \frac{1}{4}r^4\sin[2](\theta) + \hbar r \,, \nn \\
&=:& \frac{1}{2}p_r^2 + \frac{2}{r^2}p_\theta^2 + V_{eff} \,.
\label{Hamiltoniannewcoord}
\eeqa

\begin{figure}[!htb]
	\begin{subfigure}[!htb]{.5\textwidth}
		\centering
		\includegraphics[width=6.5cm]{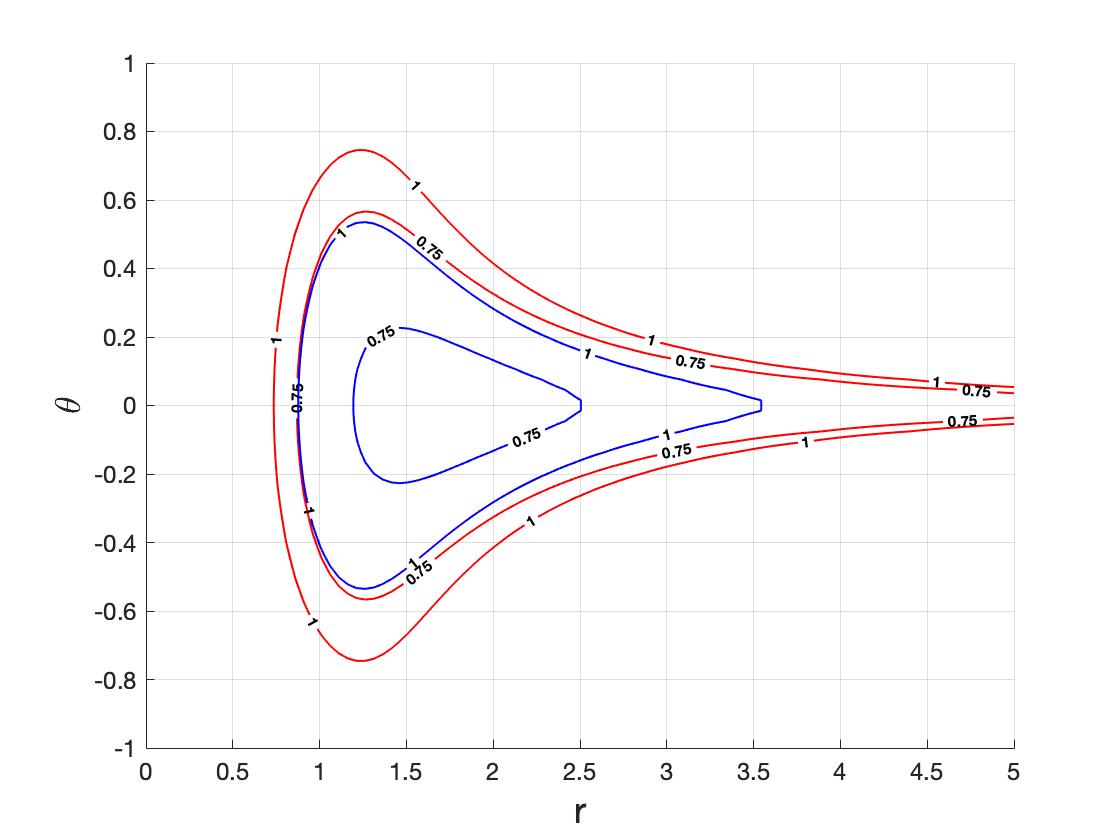}
		\caption{$p_\phi=0$ and $\kappa =0$ (red lines) , $ 4 \pi \kappa = \pm 3$ (blue lines)}
		\label{fig:ctrpf0fig1}
	\end{subfigure}
	\begin{subfigure}[!htb]{.5\textwidth}
	\centering
	\includegraphics[width=6.5cm]{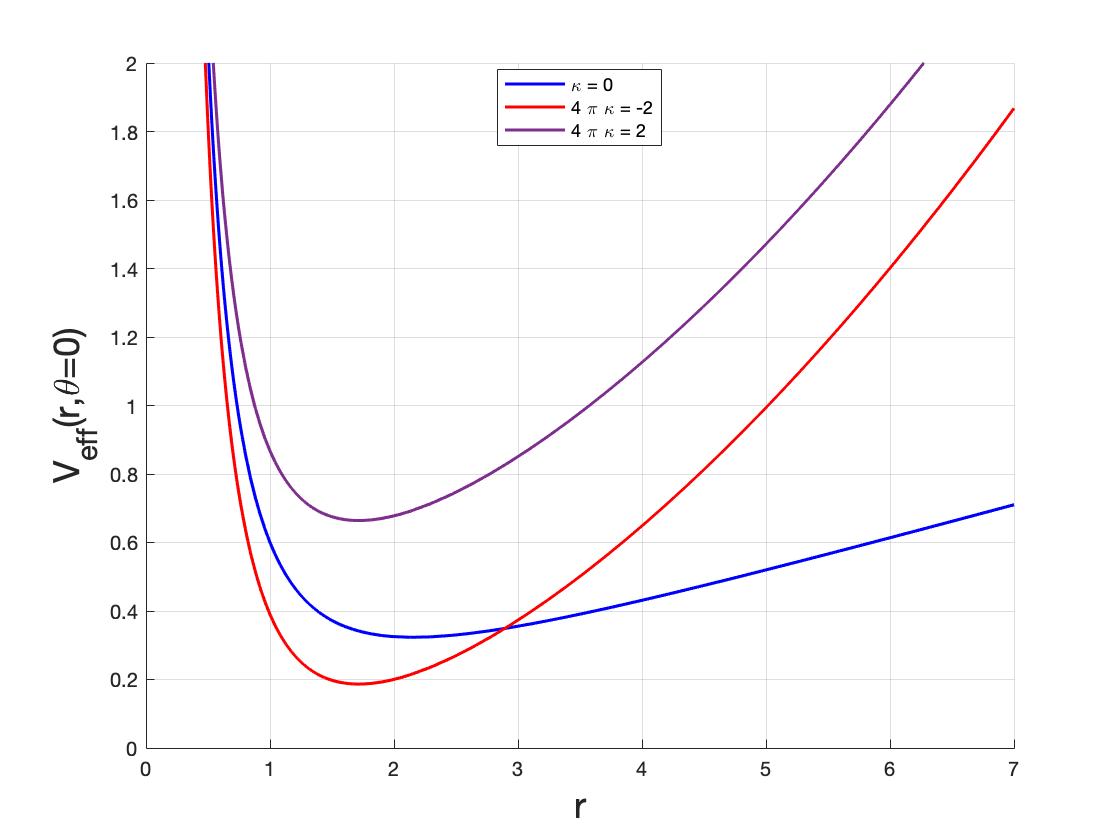}
	\caption{$p_\phi=1$ and $V_{eff}(r, \theta=0)$}
	\label{fig:veffk2rv2}
\end{subfigure}
\begin{subfigure}[!htb]{.5\textwidth}
		\centering
		\includegraphics[width=6.5cm]{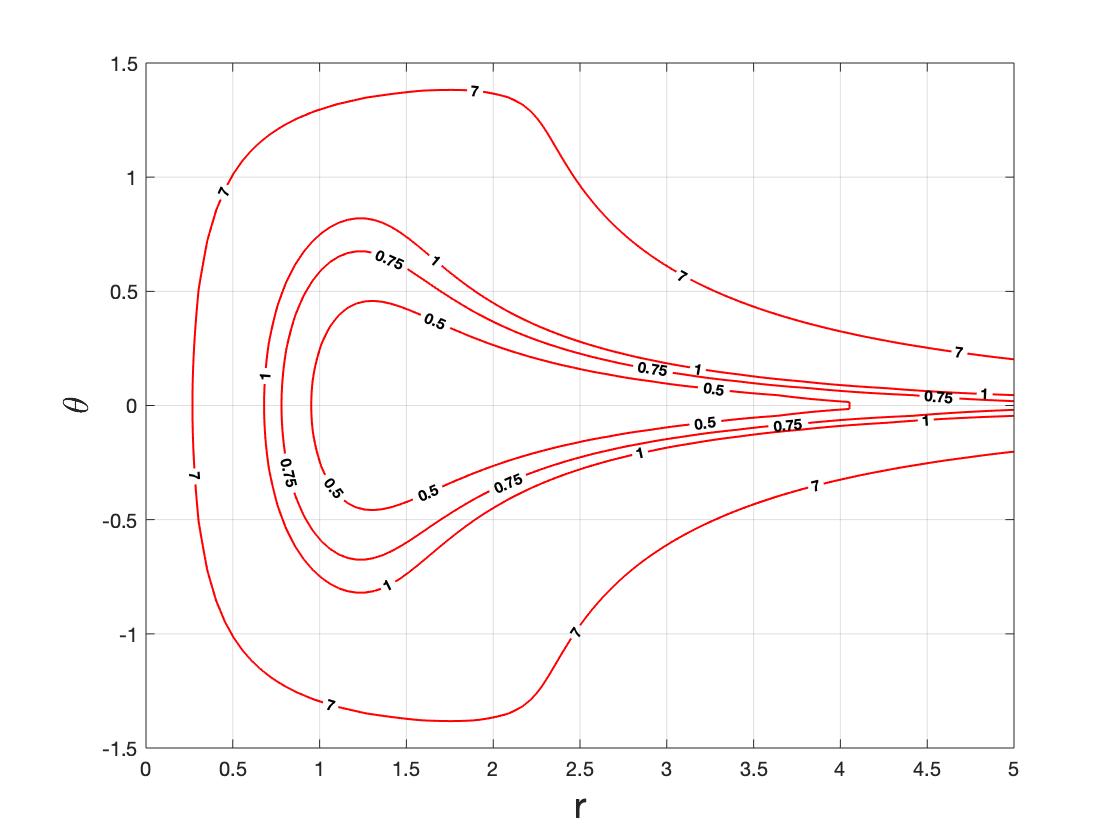}
		\caption{$p_\phi=1$ and $4 \pi \kappa = -2$}
		\label{fig:ctrpf1fig1k-2}
	\end{subfigure}
\begin{subfigure}[!htb]{.5\textwidth}
	\centering
	\includegraphics[width=6.5cm]{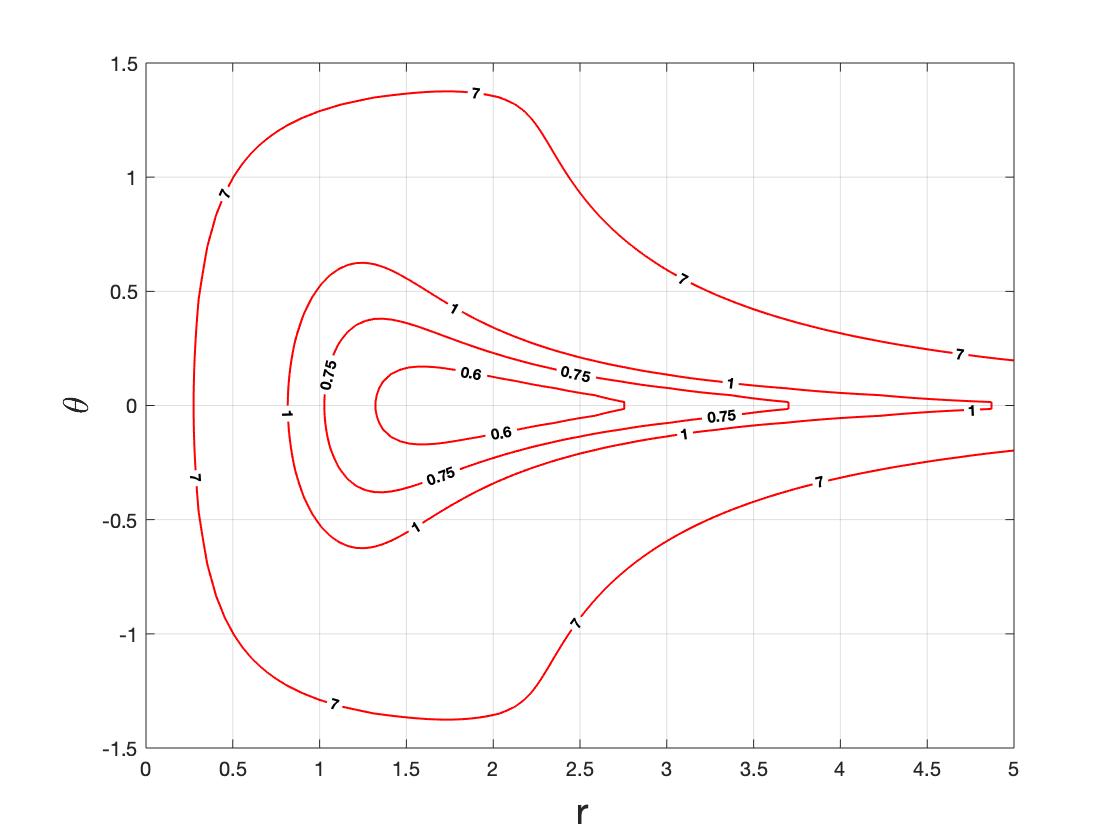}
	\caption{$p_\phi=1$ and $4 \pi \kappa = 2 $}
	\label{fig:ctrpf1fig2k2}
\end{subfigure}
\caption{Contour plots for $V_{eff} $ in $(a), (c)$, \& $(d)$ and $V_{eff}(r, \theta=0)$ in $(b)$.}
\label{fig:ctrfigs}
\end{figure}

Since this Hamiltonian is cyclic in $\phi$, as in the pure YM case \cite{Berenstein:2016zgj}, $p_\phi$ is a  constant of motion and taking advantage of this fact, we have defined the effective potential, $V_{eff}$, in the second line of (\ref{Hamiltoniannewcoord}). A number of remarks regarding this Hamiltonian are now in order\footnote{For the pure Chern-Simons limit of this Hamiltonian, readers are referred to the appendices $C$ and $D$.}. Firstly, we observe that the terms involving the CS coupling $\kappa$ are new and therefore we are now in a position to examine the chaotic dynamics emerging from (\ref{Hamiltoniannewcoord}) as $\kappa$ and the angular momentum $p_\phi$ assume a range of different values. Also note the presence of the $\hbar r$ term in $V_{eff}$. In \cite{Berenstein:2016zgj} this term is motivated by the fact that for $\sin \theta \approx \theta$, the motion can be considered to be adiabatic in $\theta$ with an effective frequency $\omega_{\theta, eff} \approx r$. With $\hbar$ taken as a small parameter, the term  $\hbar r$ can then be considered as the quantum mechanical correction to the energy, which lifts the flat direction of the pure YM model, that is, the case corresponding to the commuting matrices.  In the present case, dependence of $V_{eff}$ on $\theta$ is the same as the pure YM model, leading to the same interpretation for this term.  Though, the interesting new fact is that, for $\kappa \neq 0$, $V_{eff}$ already develops a minimum even at $\hbar =0$. This minimum is at $\theta =0$, and the real positive root of the quartic equation $\kappa^2 r^4 + \hbar r^3  - p_\phi^2 =0$. For $\hbar =0$, we obtain $r^2 = \left | \frac{p_\phi}{\kappa} \right |$, which yields $E > 2 \kappa  p_\phi $ for $\kappa p_\phi  > 0$ and simply  $E > 0 $ for $\kappa p_\phi  < 0$. Let us also note that for $\kappa=0$, $ r \propto p_{\phi}^{2 / 3} \hbar^{-1 / 3}$, and for a typical value of $\hbar =0.1$, $V_{eff} \approx 0.32$ at $p_\phi =1$ \cite{Berenstein:2016zgj}, while for $\kappa \neq 0$, this minimum shifts upward for $\kappa>0$ and downward for $\kappa < 0$. For instance, we have  $V_{eff} \approx 0.53$ and $0.21$  at $4  \pi \kappa =2$ and $4 \pi \kappa =-2$, respectively; this is illustrated in figure \ref{fig:veffk2rv2}. In general, the positive shift of the $V_{eff}$ with increasing values of $\kappa  p_\phi > 0 $ reinforces the harmonic term in the potential and they together act to decrease the Lyapunov exponent, while $\kappa  p_\phi < 0 $ gives a window of negative values ($- 5 < 4 \pi \kappa < 0$),  in which we observe a slight increase in the Lyapunov spectrum as will be made manifestly in the next section. 

It is also useful to have the contour plots of the $V_{eff}$ at $p_\phi=0,1,2$ for various values of $\kappa$ as will refer to them in the next section. These are given in figure  \ref{fig:ctrpf0fig1}, \ref{fig:ctrpf1fig1k-2} and \ref{fig:ctrpf1fig2k2}. Sharp edges in these potential contours near $\theta \approx 0$ correspond to the flat direction of the pure YM potential. In the present case, the CS term helps to lift this, as the harmonic term in  $V_{eff}$ assists to shrink the sharp edges for all values of $p_\phi$ and also acts to pull the contours toward closed loops for $p_\phi \neq 0$. For $p_\phi  > 0$, the latter happens faster for $\kappa > 0$ as opposed  to $ \kappa < 0$ and vice versa for $p_\phi < 0$.

\section{Analysis of the Chaotic Dynamics}

We now explore the chaotic structure of the system governed by (\ref{Hamiltoniannewcoord}) by studying the Lyapunov spectrum and the Poincar\'{e} sections.

\subsection{Lyapunov Spectrum}

\begin{figure}[!htb]
	\begin{subfigure}[!htb]{.5\textwidth}
		\centering
		\includegraphics[width=7.9cm]{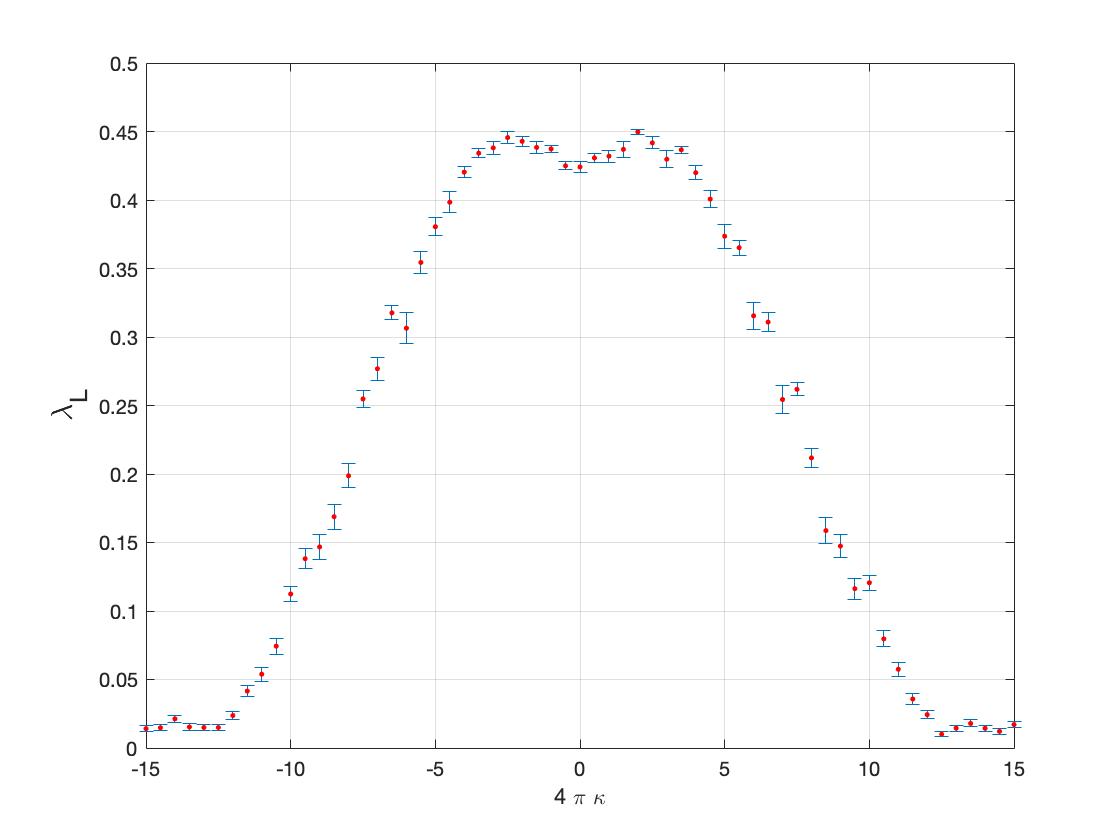}
		\caption{$p_\phi=0$}
		\label{fig:pf0fig1v2}
	\end{subfigure}
	\begin{subfigure}[!htb]{.5\textwidth}
		\centering
		\includegraphics[width=7.9cm]{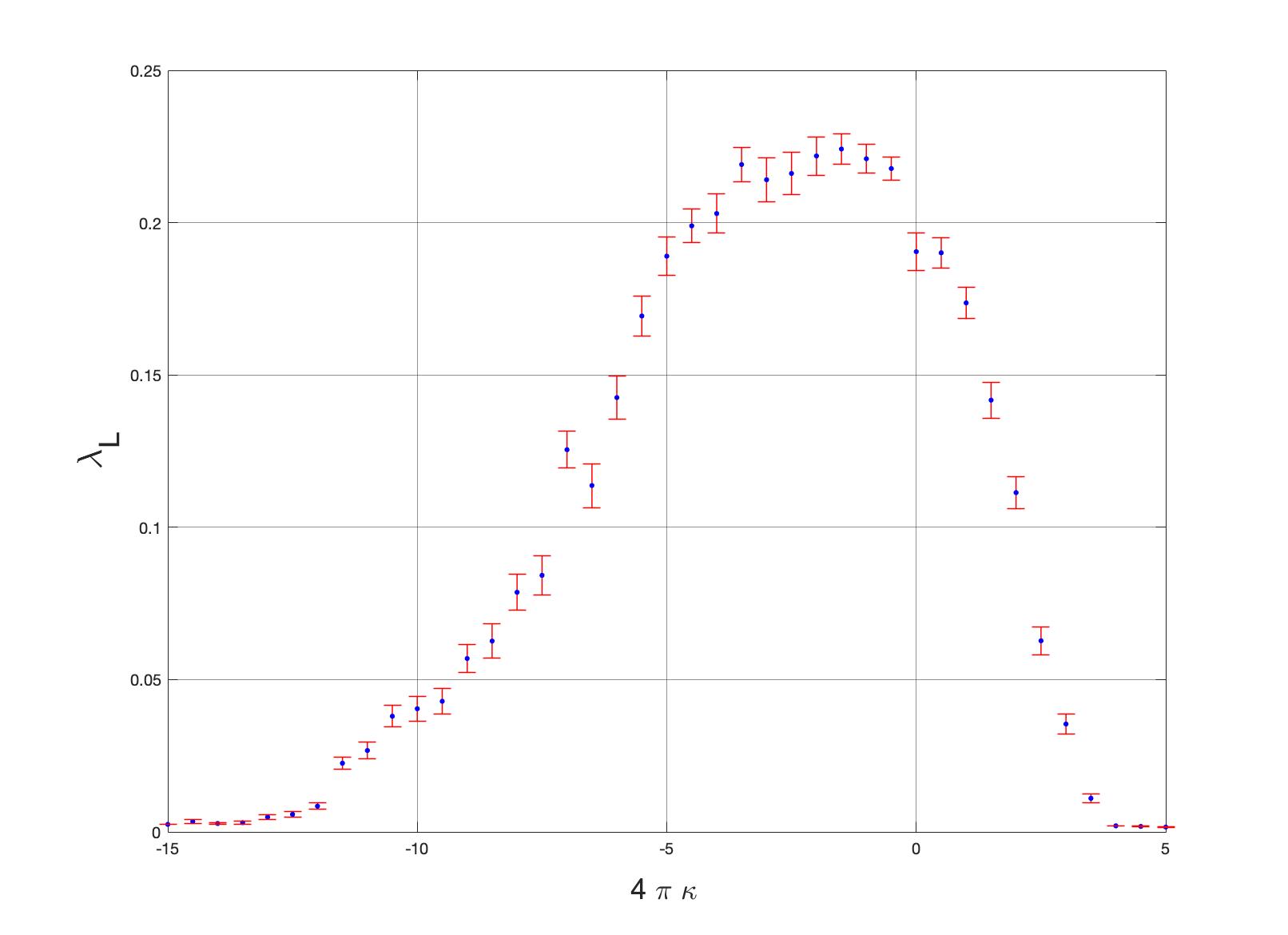}
		\caption{$p_\phi=1$}
		\label{fig:pf1fig1v2}
	\end{subfigure}
	\begin{subfigure}[!htb]{\textwidth}
		\centering
		\includegraphics[width=7.9cm]{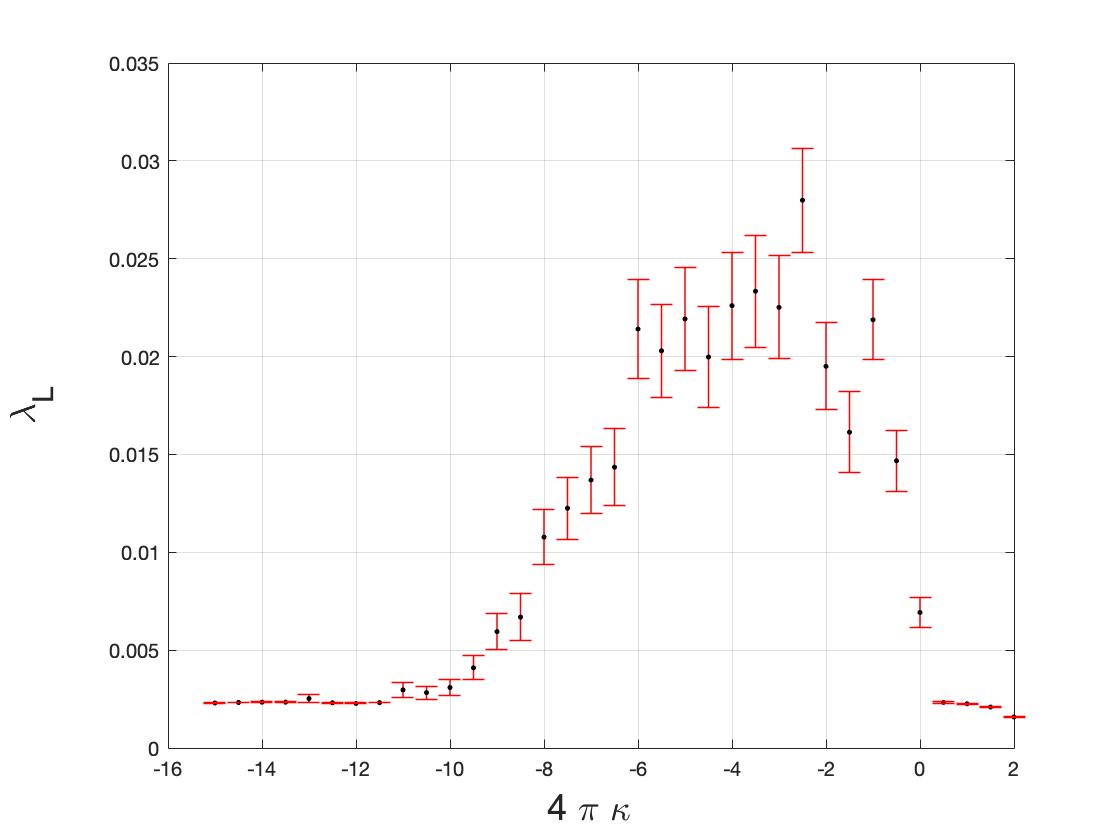}
		\caption{$p_\phi=2$}
		\label{fig:pf2fig1v2}
	\end{subfigure}
	\caption{Lyapunov spectra versus $\kappa$ values at $p_\phi=0,1,2$.}
	\label{fig:lpf}
\end{figure}

Setting the energy $E =1$, $\hbar =0.1$, and letting $p_\phi$ assume the values $0,1,2$, which is convenient for ease in comparison with the pure YM  matrix model results in \cite{Berenstein:2016zgj}, we obtain the Largest  Lyapunov Exponents (LLE), $\lambda_L$, as the CS coupling takes on a range of values, in which the typical behavior of the LLE's are captured. Our results are obtained after averaging over $120$ randomly selected initial conditions\footnote{Our method for evaluating the Lyapunov exponents and choosing initial conditions are explained in the appendix \ref{IC}.} in each case. They are presented in the figures \ref{fig:pf0fig1v2},  \ref{fig:pf1fig1v2}, \ref{fig:pf2fig1v2} and we will elaborate on them shortly. 

Chaotic structure of the pure YM model is explored in \cite{Berenstein:2016zgj} and it is found that the system is fully chaotic at $p_\phi=0$ and essentially becomes non-chaotic with increasing values of  $p_\phi$.  At the intermediate values $0 < p_{\phi} < 2$, for example at $p_{\phi} =1$, there are regions in the phase space, in which quasi-periodic motion is present as signaled by KAM tori appearing in the Poincar\'{e} section plots given in \cite{Berenstein:2016zgj}, while the rest of the phase space is filled with chaotic motion.  

In figure \ref{fig:pf0fig1v2}, profile of the Lyapunov spectrum of the model values of $\kappa$ in the interval $ 4 \pi |\kappa| \leq 15 $ and at $p_\phi=0$ is presented. The plot is essentially symmetric w.r.t. the $\kappa = 0$-axis as may be expected from (\ref{Hamiltoniannewcoord}), which is even under $\kappa \leftrightarrow - \kappa$ for $p_{\phi} = 0$ and although LLE values tend to decrease in an almost monotonic manner for $4 \pi \abs{\kappa}>4$, they are essentially non-vanishing for $ 4 \pi |\kappa| \leq 10$, which makes us conclude that the model is chaotic and behaves similar to the pure YM case within this range of the CS coupling.  The rather mild increase in the LLE values observed in this plot in the narrow range $ 4 \pi \abs{\kappa} < 4$ can be explained as follows.  As $\abs{\kappa}$ increases, sharp edged regions in the contour plot of the effective potential $V_{eff}$, as illustrated in figure \ref{fig:ctrpf0fig1},  become less pronounced, and consequently, compared to $\kappa =0$, the system spends relatively less time in these regions where the dynamics is adiabatic in $\theta$ and therefore no appreciable contribution to chaos is expected \cite{Berenstein:2016zgj}. In turn, this results in a mild increase in the LLE spectrum within the indicated range of $\kappa$ values. Nevertheless, for $  4 \pi |\kappa| > 4$,  harmonic term starts to become significant and the chaotic dynamics is gradually lost. 

For $p_\phi \neq 0$,  the $\kappa p_\phi$ term in  $V_{eff}$ impacts the Lyapunov spectrum asymmetrically depending on its sign, as it causes a fixed negative or a positive shift on the latter. At  $p_\phi =1$, for instance, which is illustrated in figure \ref{fig:pf1fig1v2}, we immediately observe that $\lambda_L$ values with in the range of values $ -5 <  4 \pi \kappa  < 0$ are above what is computed at $\kappa = 0$. This can be attributed to the downward shift in $V_{eff}$ due to  $\kappa p_\phi < 0 $, which clearly also lowers the minimum of $V_{eff}$ as we have already discussed toward the end of  previous section. The increase in $\lambda_L$ can not be sustained for $  4 \pi \kappa  < -5$, since then the harmonic term $\propto \kappa^2 r^2$ becomes sufficiently strong even at short distances to dominate $V_{eff}$ and initiates the decline of the chaotic dynamics. For $\kappa > 0$,  this term acts to strengthen the harmonic terms and the chaotic motion becomes sharply suppressed before $ 4 \pi \kappa  \approx 5$. At  $p_\phi =2$, we still observe a mild increase in the Lyapunov exponents roughly in the range $-4 <  4 \pi \kappa  < 0$, but the maximum value of $\lambda_L$ now appears to be $\approx 0.03$, an order of magnitude less than that is found for $p_\phi =0$ and $p_\phi =1$, and not significant enough to conclude that any dense chaotic dynamics remain for $p_\phi \geq 2$. 

\subsection{Poincar\'{e} sections}

\begin{figure}[!htb]
	\begin{subfigure}[!htb]{.32\textwidth}
		\centering
		\includegraphics[width=5.75cm]{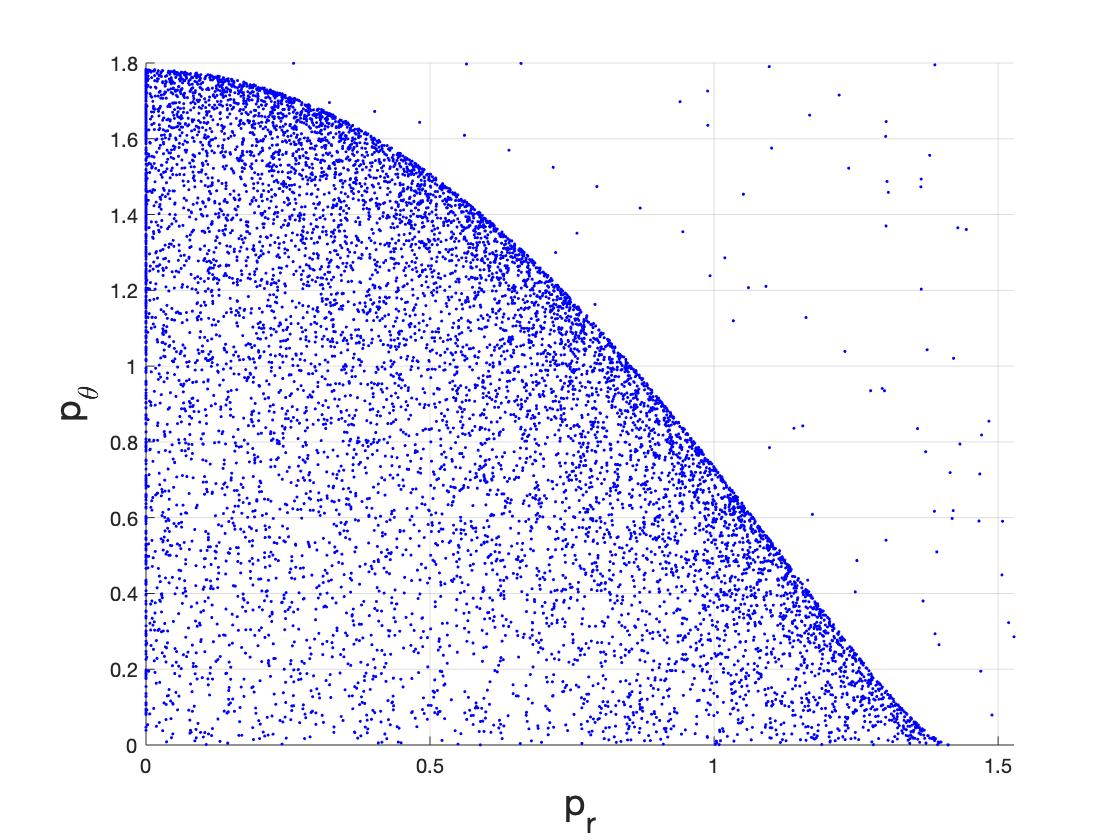}
		\caption{$  4 \pi \kappa = \pm 2$}
		\label{fig:pf0kpm2}
	\end{subfigure}
	\begin{subfigure}[!htb]{.32\textwidth}
		\centering
		\includegraphics[width=5.75cm]{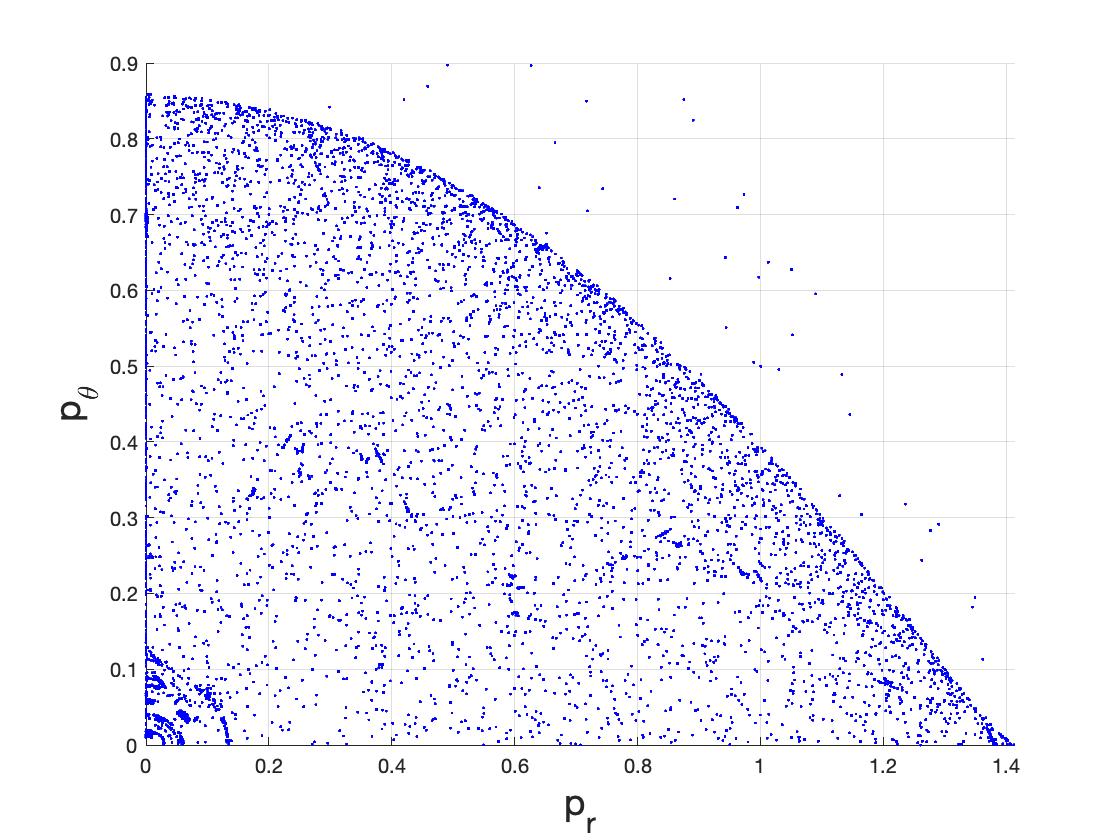}
		\caption{$  4 \pi \kappa  = \pm 6$}
		\label{fig:pf0kpm6}
	\end{subfigure}
\begin{subfigure}[!htb]{.3\textwidth}
	\centering
	\includegraphics[width=5.75cm]{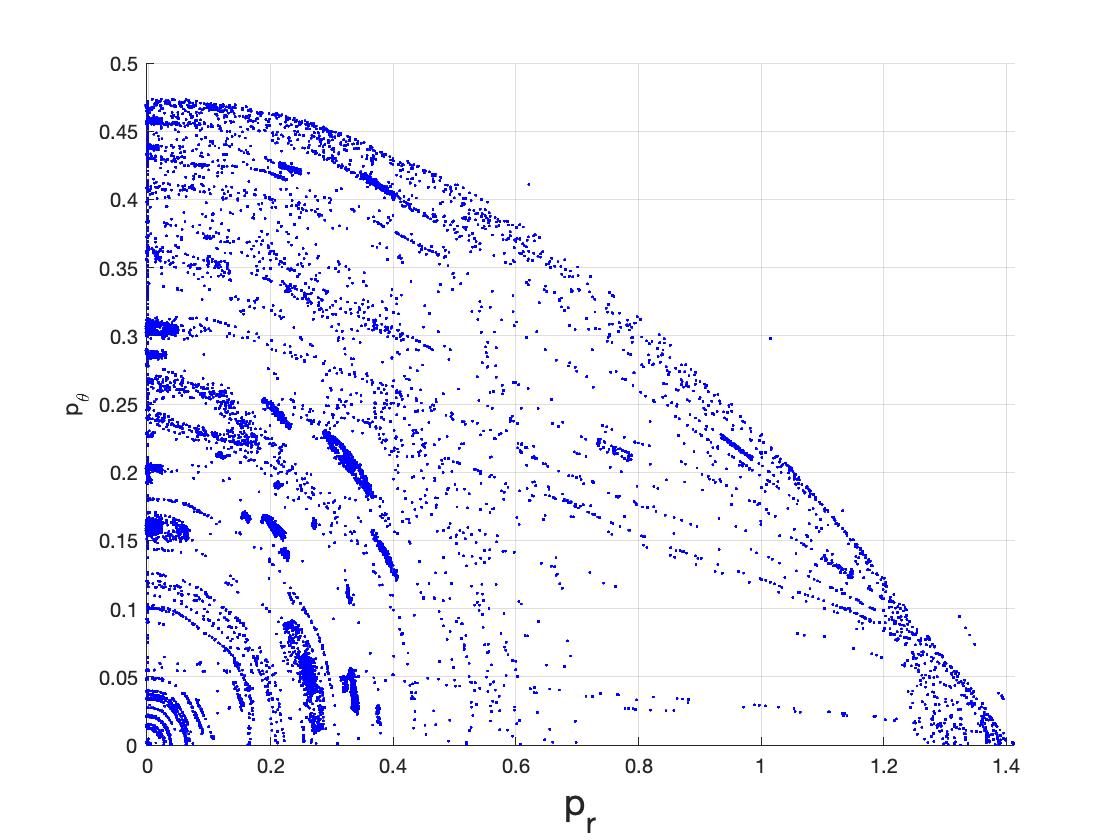}
	\caption{$  4 \pi \kappa  = \pm 12$}
	\label{fig:pf0kpm12}
\end{subfigure}
	\caption{Poincar\'{e} sections at $p_\phi= 0$}
	\label{fig:poincare_cs_kppf0}
\end{figure}

\begin{figure}[!htb]
	\begin{subfigure}[!htb]{.3\textwidth}
		\centering
		\includegraphics[width=5.75cm]{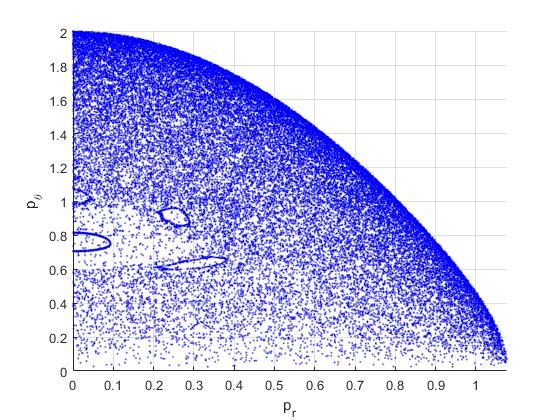}
		\caption{$  4 \pi \kappa =1$}
		\label{fig:pf1k1}
	\end{subfigure}
	\begin{subfigure}[!htb]{.3\textwidth}
		\centering
		\includegraphics[width=5.75cm]{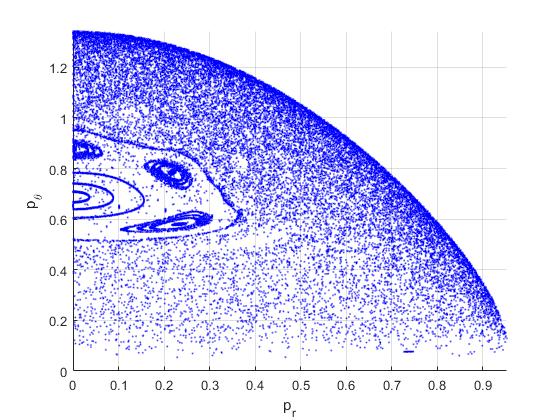}
		\caption{$4 \pi \kappa=2$}
		\label{fig:pf1k2}
	\end{subfigure}
	\begin{subfigure}[!htb]{.3\textwidth}
		\centering
		\includegraphics[width=5.75cm]{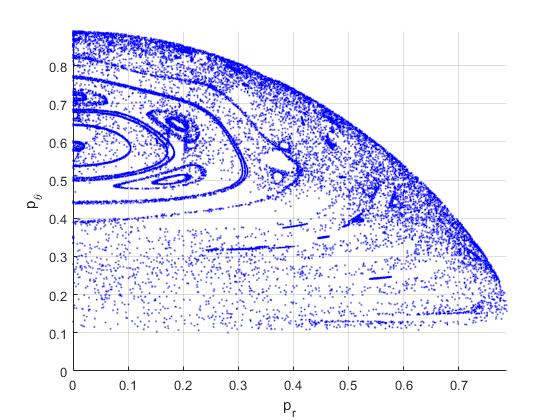}
		\caption{$4 \pi \kappa=3$}
		\label{fig:pf1k3}
	\end{subfigure}
	\begin{subfigure}[!htb]{\textwidth}
		\centering
		\includegraphics[width=5.75cm]{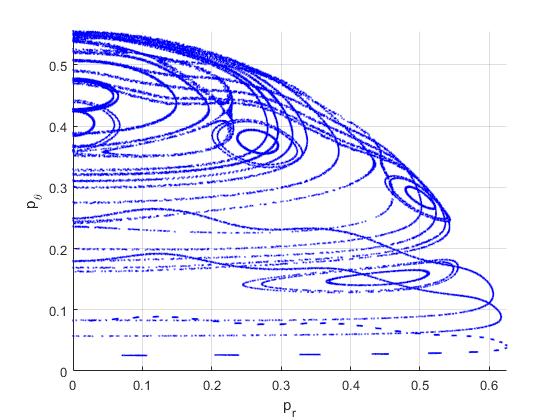}
		\caption{$4 \pi \kappa=4$}
		\label{fig:pf1k4}
	\end{subfigure}
	\caption{Poincar\'{e} sections at $p_\phi=1$ for $\kappa > 0$}
	\label{fig:poincare_cs_kppf1}
\end{figure}

\begin{figure}[!htb]
	\begin{subfigure}[!htb]{.3\textwidth}
		\centering
		\includegraphics[width=5.75cm]{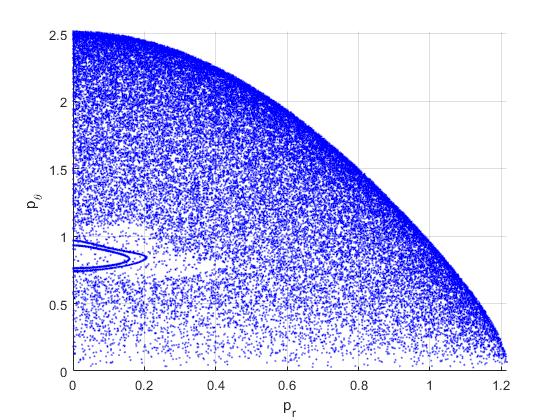}
		\caption{$ 4 \pi \kappa =-1$}
		\label{fig:pf1km1}
	\end{subfigure}
	\begin{subfigure}[!htb]{.3\textwidth}
		\centering
		\includegraphics[width=5.75cm]{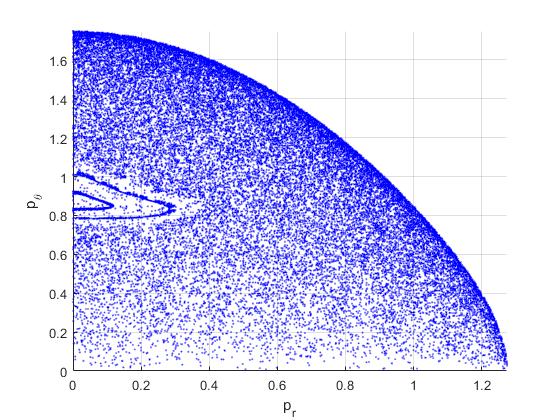}
		\caption{$ 4 \pi \kappa=-3$}
		\label{fig:pf1km3}
	\end{subfigure}
	\begin{subfigure}[!htb]{.3\textwidth}
		\centering
		\includegraphics[width=5.75cm]{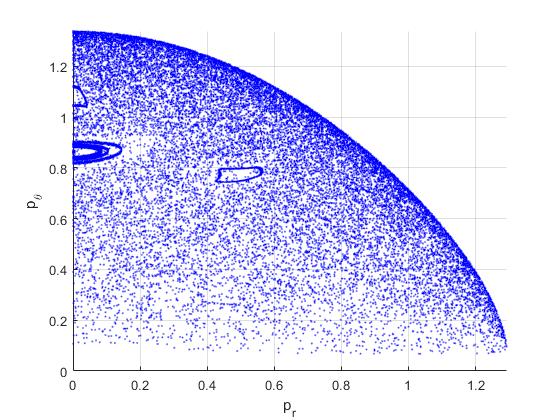}
		\caption{$ 4 \pi \kappa =-5$}
		\label{fig:pf1km5}
	\end{subfigure}
	\begin{subfigure}[!htb]{.33\textwidth}
	\centering
	\includegraphics[width=5.75cm]{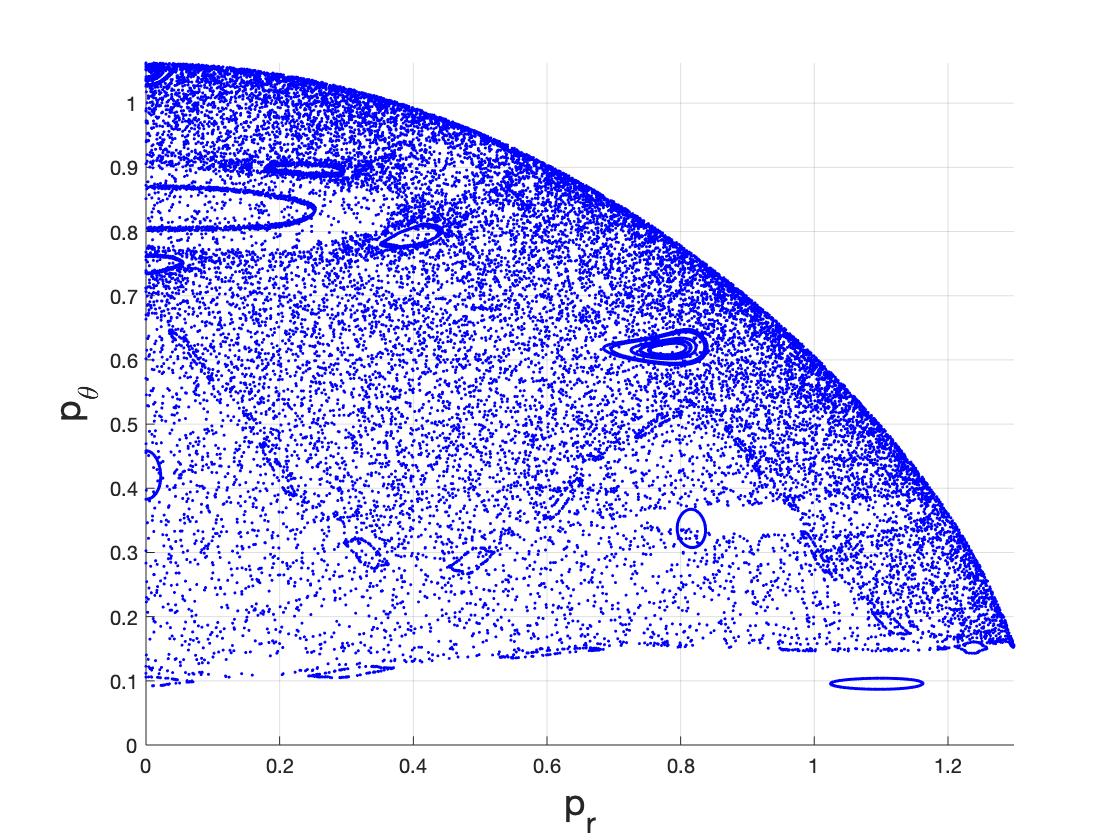}
	\caption{$4 \pi \kappa =-7.5$}
	\label{fig:pf1km75}
\end{subfigure}
	\begin{subfigure}[!htb]{.32\textwidth}
		\centering 
		\includegraphics[width=5.75cm]{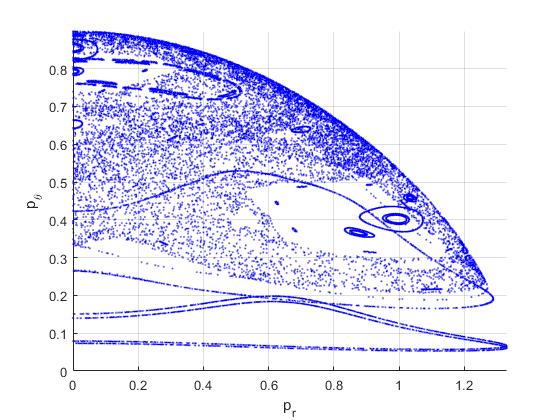}
		\caption{$4 \pi \kappa =-10$}
		\label{fig:pf1km10}
	\end{subfigure}
	\begin{subfigure}[!htb]{.32\textwidth}
		\centering 
		\includegraphics[width=5.75cm]{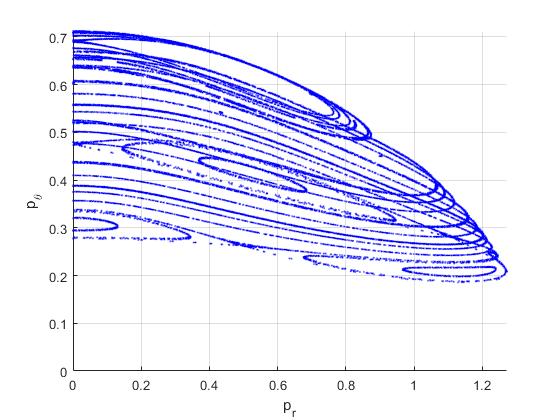}
		\caption{$4 \pi \kappa =-15$}
		\label{fig:pf1km15}
	\end{subfigure}
	\caption{Poincar\'{e} sections at $p_\phi=1$ for $\kappa < 0 $}
	\label{fig:poincare_cs_kmpf1}
\end{figure}

\begin{figure}[!htb]
	\begin{subfigure}[!htb]{.3\textwidth}
		\centering
		\includegraphics[width=5.75cm]{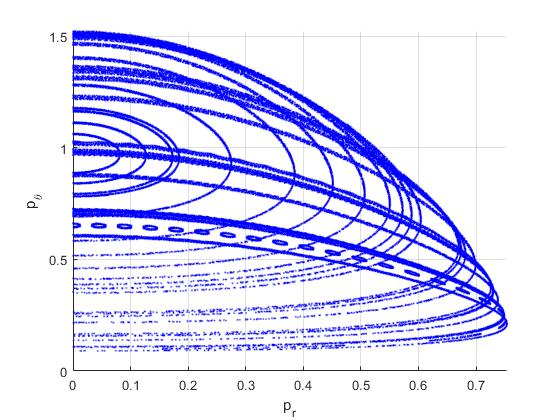}
		\caption{$ 4 \pi \kappa =1$}
		\label{fig:pf2k1}
	\end{subfigure}
	\begin{subfigure}[!htb]{.3\textwidth}
		\centering
		\includegraphics[width=5.75cm]{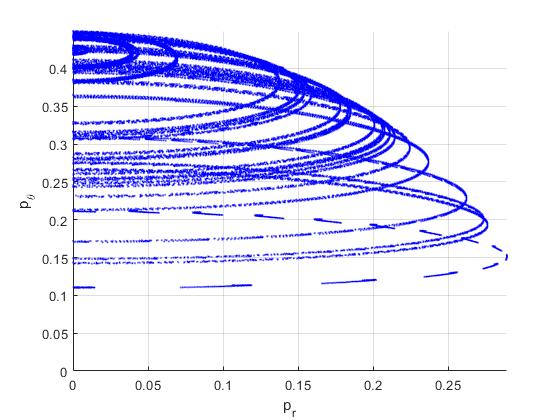}
		\caption{$ 4 \pi \kappa =2$}
		\label{fig:pf2k2}
	\end{subfigure}
	\begin{subfigure}[!htb]{.3\textwidth}
		\centering
		\includegraphics[width=5.75cm]{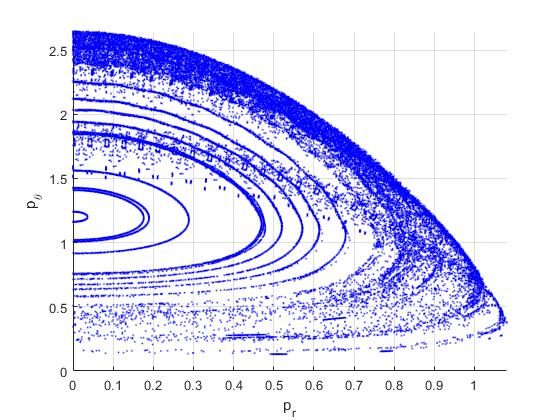}
		\caption{$4 \pi \kappa =-1$}
		\label{fig:pf2km1}
	\end{subfigure}
	\begin{subfigure}[!htb]{.3\textwidth}
		\centering
		\includegraphics[width=5.75cm]{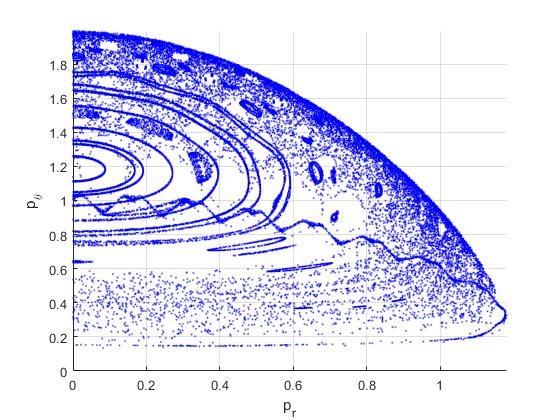}
		\caption{$4 \pi \kappa =-3$}
		\label{fig:pf2km3}
	\end{subfigure}
	\hfill
	\begin{subfigure}[!htb]{.3\textwidth}
		\centering
		\includegraphics[width=5.75cm]{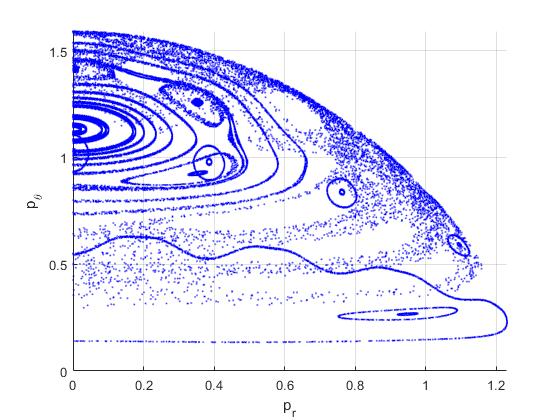}
		\caption{$ 4 \pi \kappa =-5$}
		\label{fig:pf2km5}
	\end{subfigure}
	\hfill
	\begin{subfigure}[!htb]{.3\textwidth}
		\centering
		\includegraphics[width=5.75cm]{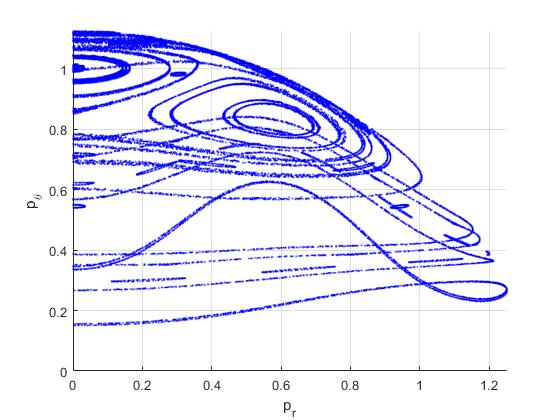}
		\caption{$ 4 \pi \kappa =-10$}
		\label{fig:pf2km10}
	\end{subfigure}
	\caption{Poincar\'{e} sections at $p_\phi=2$}
	\label{fig:poincare_cs_pf2}
\end{figure}

All of the conclusions of the previous subsection regarding the chaotic dynamics of the present dynamical system are well supported by the Poincar\'{e} sections. We have obtained the latter at the $\theta=0$ intersections of the phase space and projected on to the $p_\theta , p_r$ plane. Figures \ref{fig:poincare_cs_kppf0}, \ref{fig:poincare_cs_kppf1}, \ref{fig:poincare_cs_kmpf1} and\ref{fig:poincare_cs_pf2} show the Poincar\'{e} sections on the first quadrant of the $p_\theta$, $p_r$ plane.

From figure \ref{fig:poincare_cs_kppf0}, we see that chaotic dynamics appears to fill the phase space at $p_\phi = 0$, for a large range of values of $\kappa$, which is approximately $  4 \pi |\kappa|  \lesssim 10$, while the periodic motion starts to compete and take over after this range of $\kappa$ values as can be observed from figure \ref{fig:pf0kpm12}. 

At $p_\phi =1$ and $ 4 \pi \kappa  =1$, from figure \ref{fig:poincare_cs_kppf1}, we observe that the phase space is still dominated by chaos, while a few KAM tori indicating quasi-periodic motion are visible. As $\kappa$ continues to increase, more KAM tori start to occur, and system swiftly becomes non-chaotic for $  4 \pi \kappa  \gtrsim 4$ and gets  dominated by quasi-periodic orbits. However, for $\kappa < 0 $ as illustrated in figure \ref{fig:poincare_cs_kmpf1}, the system appears to remain densely chaotic with only a few KAM tori appearing until around $  4 \pi \kappa   \approx -5$, while the quasi-periodic motion starts to spread for $  4 \pi \kappa  \lesssim -7.5$ and start to take over only after $ 4 \pi \kappa  \lesssim -10$. Let us also note that, some KAM tori appear to intersect, especially as seen in the Poincar\'{e} sections at larger values of $|\kappa|$, for instance, in the figures (\ref{fig:pf1k4}) and (\ref{fig:pf1km15}). This is due to possible different values of the $r$ coordinate appearing in the evolution of the system starting with distinct initial conditions being projected to the same point on the $p_\theta$,$p_r$ plane.

For $p_\phi =2$, we see that there is very little chaos remaining in the phase space regardless of the value of $\kappa$ and quasi periodic motion dominates the phase space. This can be seen from the Poincar\'{e} sections in figure \ref{fig:poincare_cs_pf2}. There is no chaos for $\kappa > 0$, and although some randomly spread points appear for negative $\kappa$ values, for small $|\kappa|$, KAM tori quickly dominate the phase space and quasi periodic motion is all that is left. 

\subsection{Transition from Chaotic to Non-Chaotic Phase}

\begin{figure}[!htb]
	\centering
	\includegraphics[width=0.6\textwidth]{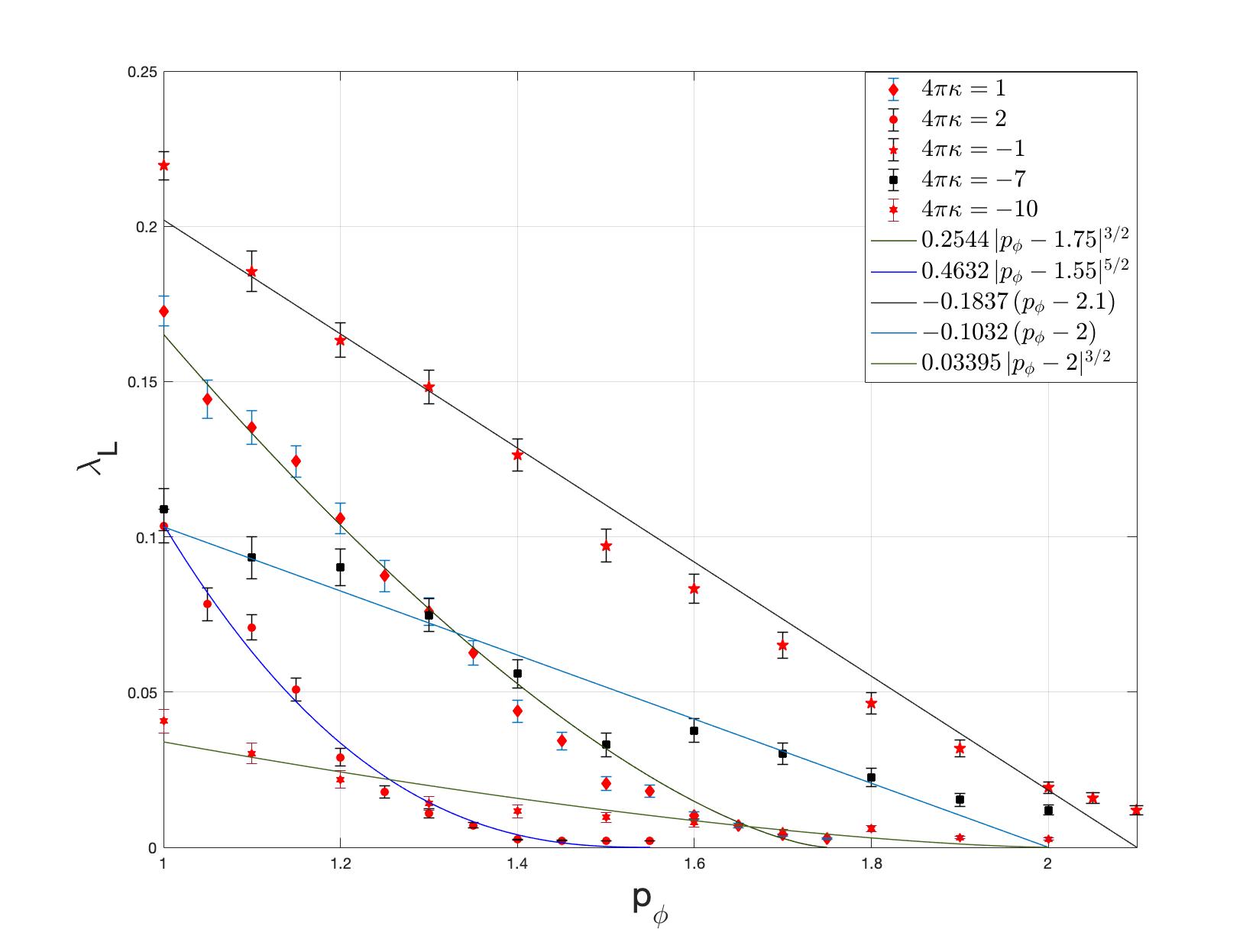}
	\caption{Lyapunov spectra versus $p_\phi$ at $4 \pi \kappa =-10,-7,-1,1,2$.  Estimates for $p_\phi^c$ and the critical exponents for $\lambda_L$ are obtained from the best fitting curves to the data. }
	\label{fig:criticalexp}
\end{figure}

In order to investigate the transition of the system from the chaotic phase, i.e. black-brane phase, to the non-chaotic, integrable phase dominated by quasi periodic motion, it is useful to examine, the change of $\lambda_L$ with $p_\phi$ treated as the order parameter, while keeping $\kappa$ fixed. The fitting curves presented in figure \ref{fig:criticalexp} helps to illustrate the situation with sufficient clarity. In particular, from this figure, we see that for $-7 \lesssim 4 \pi \kappa < 0$, $\lambda_L$ decreases linearly with increasing $p_\phi$, and the transition between the two-phases occur at $p_\phi^c \approx 2$ with a critical exponent of $1$ for $\lambda_L$.  Therefore, we conclude that approximately within this range of $\kappa$, the transition from the chaotic to non-chaotic phase has the same characteristic as found for the pure YM model in \cite{Berenstein:2016zgj}. For $4 \pi \kappa < -7$, $\lambda_L$ value is already below $0.1$ (see figure \ref{fig:pf1fig1v2})  at $p_\phi =1$ and tends to decrease faster with increasing $p_\phi$; at $4 \pi \kappa =-10$ we estimate that what little remains of the chaotic phase approaches $p_\phi^c \approx 2$ with a critical exponent $\approx 3/2$.  For $\kappa >0$, on the other hand, not only the approach to non-chaotic phase appears to be faster, but also it tends to occur at smaller values of $p_\phi$ at larger $\kappa$. For instance, we estimate that at $4 \pi \kappa =1$, $p_\phi^c \approx 1.75$, while at $4 \pi \kappa =2$, $p_\phi^c \approx 1.55$, with critical exponents $\approx3/2$ and $\approx 5/2$, respectively. 

\section{Conclusions and Outlook}

In this paper, we have studied the chaotic structure of the minimal Yang Mills Chern Simons matrix model. Using the gauge and global symmetries, and with a suitable choice of the coordinates, the Hamiltonian of the system is obtained in a form in which the Gauss law constraint is fully solved and manifestly imposed. We have studied the chaotic dynamics of the model, and in particular, probed the changes in the Lyapunov exponent as the values of both the CS coupling, $\kappa$, and the conserved conjugate momentum, $p_\phi$, are varied. We have found that, even for $p_\phi=0$, there is a range of CS coupling values, approximately given as $4 \pi |\kappa| \lesssim 4$  within which the Lyapunov exponent is larger in value compared to that evaluated at $\kappa=0$. We have also seen that $\kappa p_\phi$ term in the effective potential alters the Lyapunov spectrum depending on its sign.  We have found that the largest Lyapunov exponents evaluated within a range of values of $\kappa$ are above that is computed at $\kappa=0$, for $\kappa p_\phi < 0$. These results are discussed in detail in section 3, where we also presented estimates for the critical exponents for $\lambda_L$ and the value of the order parameter, $p_\phi^c$, as the system transits from chaotic to non-chaotic phase.

Let us finally note that out of time order correlators (OTOCs) approach recently applied in \cite{Akutagawa:2020qbj} to a system involving a $x^2 y^2$ term in the potential to probe quantum chaos may also be suitable for the model treated in this paper and we hope to report on any developments along these directions elsewhere in the near future.




\vskip 1em

\noindent{\bf \large Acknowledgments}

\vskip 1em

\noindent Part of S.K.'s work was carried out during his sabbatical stay at the physics department of CCNY of CUNY and he thanks V.P. Nair and D. Karabali for the warm hospitality at CCNY and the metropolitan area. S.K. thanks A.P.Balachandran for discussions and critical comments. Authors acknowledge the support of TUBİTAK under the project number 118F100 and the METU research project GAP-105-2018-2809. 

\vskip 1em


\appendices

\section{Chern-Simons Action in $0+1$-Dimensions}\label{DR}

In $2+1$-dimensions the CS action for the Hermitian $SU(2)$ gauge fields $A_\mu$ may be given as \cite{Dunne},\cite{Bal}
\be
S_{CS \, (2+1)} = \frac{k}{4 \pi} \, \int \, d^3 x \, \epsilon^{\mu\nu\rho} \Tr(- A_\mu \partial_{\nu} A_\rho + \frac{2}{3}iA_\mu A_\nu A_\rho) \,,
\ee
where $A_\mu$ transform under the $SU(2)$ gauge transformations as $A_\mu \rightarrow U^\dagger A_\mu U + i U^\dagger \partial_\mu U$. Invariance of 
$e^{i S_{CS \, (2+1)}}$ under large gauge transformations requires that $k$ is an integer\footnote{Let us also note that for gauge fields valued in the Lie algebra of the Lie group $G$, CS action on a $2D+1$ dimensional space-time manifold $M$ changes under $G$ transformation by a term of the form $\Omega \propto \int_M  \, Tr \lbrack  (g^{-1} d g)^{2D+1} \rbrack$ (up to a constant factor) where $g \in G$. In general $\Omega$ does not vanish \cite{Bal}. For $M = S^{2D+1}$, $\Omega$ is determined via the $(2D+1)^{th}$ homotopy group of $G$, $\pi_{2D+1}(G)$, and the requirement of gauge invariance of $e^{i S_{CS}}$ leads to the level quantization of the CS coupling. If $M$ is the $2D+1$ Minkowski space, $\Omega$ is still determined by $\pi_{2D+1}(G)$ if we demand that $g \rightarrow 1$ as time goes to $\pm \infty$ and at spatial boundaries \cite{Bal, Dunne}. For $D = 0$ and $G \equiv SU(2)$, we clearly have $\Omega =0$ since $g^{-1} d g$ is valued in the Lie algebra of $SU(2)$ and hence its trace is vanishing. Thus, this general consideration is applicable to the $S_{CS}$ in $0+1$-dimensions and indicates that $S_{CS \, (0+1)}$ is indeed gauge invariant.  In particular, we also have that $\pi_{1}(SU(2)) \equiv \Pi_{1}(S^3) = 0$ therefore $\Omega$ vanishes in $0+1$-dimensions implying once again that $S_{CS \, (0+1)}$ is gauge invariant.}.

Dimensional reduction from $2+1$ to $0+1$-dimension is facilitated by requiring that all spatial dependence of the gauge fields $A_\mu(x,t)$ are dropped and only their time dependence is retained. This implies that in $S_{CS \, (2+1)}$ all the spatial derivatives collapse to zero. Introducing the notation $A_\mu \equiv  (A_0 \,, X_i)$, $i:1,2$, we immediately find 
\beqa
S_{CS \, (0+1)}  &=&  \frac{k}{4 \pi} \int \, d^2 x \, \int dt \,  \Tr(\epsilon^{ij}(X_i \dot{X}_j + 2iA_0X_iX_j)) \,, \nn \\
&=& \frac{k V_2}{4 \pi} \, \int dt \, \Tr \lbrack \epsilon_{ij} X_i  \,( D_0 X_j) \rbrack  \,, \nn \\
&=:& \kappa \, \int dt \, Tr \lbrack \epsilon_{ij} X_i  \,( D_0 X_j) \rbrack 
\label{acCS1}
\eeqa
where $V_2:= \int d^2 x$ denotes the result of the $2$-dimensional volume integral. Thus the coupling of the CS action in $0+1$ dimensions take the form $\kappa := \frac{k V_2}{4 \pi}$ and due to the volume factor $V_2$ it differs from the coupling of the CS action in 2+1-dimensions. In particular, due to the $V_2$ factor, which can take arbitrary real values, $\kappa$ is not an integer multiple of $\frac{1}{4 \pi}$. We also see that the CS coupling $\kappa$ obtained in this manner is consistent with the fact that $S_{CS \, (0+1)}$ is manifestly gauge invariant (as it is trivially observed form the second line of (\ref{acCS1}) upon using the cyclicity of the trace and $U^\dagger U = U U^\dagger = 1$). These facts clearly indicate that $\kappa$ is not level quantized. 

In this paper we work with $ 4 \pi |\kappa| \lessapprox 10$ as this conveniently gives the relevant range of $\kappa$ values to explore the dynamics at $E = 1$ and $| p_\phi | \lessapprox 2$. The latter are the values of energy and $p_\phi$ used in \cite{Berenstein:2016zgj}, which we use since it gives us an ease in comparison. In our study, we are only considering the classical theory but writing out the $\hbar$ explicitly in the action there will not be any reason to keep $ 4 \pi \kappa$ within the values of $O(1)$. 

\section{Calculation of Lyapunov Exponents and Selection of Initial Conditions}\label{IC}

\subsection{Calculation of Lyapunov Exponents}

Lyapunov exponents are useful to determine the sensitivity of a system to given initial conditions. More precisely, they measure the exponential growth in perturbations and therefore give a reliable way to establish the presence of chaos in a dynamical system \cite{Ott,Hilborn,Campbell}. For a Hamiltonian system if we denote the perturbations in the phase space coordinates ${\bm g}(t) \equiv (g_1(t), g_2(t)\,,\cdots\,, g_{2N}(t))$ by $\delta {\bm g}(t)$, then we may conclude that the system is chaotic if, at large $t$, $\delta {\bm g}(t)$ deviates exponentially from its initial value at $t= t_0$: $||\delta {\bm g}(t)|| = e^{\lambda (t-t_0)} ||\delta {\bm g}(t_0)||$. Here $\lambda > 0$ are called the positive Lyapunov exponents and there are $2N$ of them for a phase space of dimension $2N$. Let us also note that this description is in parallel with the statement that even slightly different initial conditions give trajectories in the phase space, which are exponentially diverging from each other and hence lead to chaos. In a dynamical system presence of at least one positive Lyapunov exponent is sufficient to conclude the presence of chaotic motion. In Hamiltonian systems, due to the symplectic structure of the phase space, Lyapunov exponents appear in $\lambda_i$ and $-\lambda_i$ pairs and a pair of the Lyapunov exponents vanishes as there is no exponential growth in perturbations along the direction of the trajectory specified by the initial condition. and sum of all the Lyapunov exponents is zero as a consequence of Liouville's theorem. These facts are well-known and their details may be found in many of the excellent books on chaos \cite{Ott,Hilborn,Campbell}.  

We follow the appendix in \cite{Asano:2018nol} to describe the method to compute all the Lyapunov exponents.With $U(t)$ denoting a time evolution operator we may write 
\begin{equation}
\delta {\bm g}(t) = U(t) \delta {\bm g}(0) \quad i = 1,\cdots,2N \,,
\end{equation}
and
\begin{equation}
\delta{\bm g}(t_1 + t_2) = U(t_2) U(t_1) \delta {\bm g}(0) \,.
\end{equation}
Lyapunov exponents are defined by
\begin{equation}
\lambda_i=\lim_{t\to\infty}\frac{1}{t}\log(\frac{||\delta {\bm g}(t)||}{||\delta {\bm g}(0)||})\,.
\end{equation}

Dividing the time into $n$ equal steps such that $t=n\Delta t$ Lyapunov exponents can be expressed as
\begin{equation}
\lambda =\lim_{t\to\infty}\frac{1}{n\Delta t}\log(\frac{|| U(\Delta t)...U(\Delta t) \delta {\bm g}(0)||}{||\delta {\bm g}(0)||})\,.
\end{equation}

We may consider that $({\bm h}_0^1\,,{\bm h}_0^2\,, \cdots\,, {\bm h}_0^{2N})$ span an orthonormal basis for the set of vectors tangent to the phase space trajectory ${\bm g}(0)$ at $t=0$. After a time $\Delta t$, time evolved vectors can be written as $\bm{k}_1^i = U(\Delta t) {\bm h}_0^i$, where $({\bm k}_1^1\,,{\bm k}_1^2\,, \cdots\,, {\bm k}_1^{2N})$ spans a basis of tangents vectors to the trajectory  ${\bm g}(\Delta t)$. However, this basis of vectors need not be orthogonal. Using the Gram-Schmidt orthogonalization process we can obtain the orthogonal set from $({\bm k}_1^1\,,{\bm k}_1^2\,, \cdots\,, {\bm k}_1^{2N})$, which we may denote as $(\tilde{{\bm h}}_1^1\,,\tilde{{\bm h}}_1^2\,, \cdots\,, \tilde{{\bm h}}_1^{2N})$.
The expansion rate of the vector $\tilde{{\bm h}}_1^i$ can be determined as
\begin{equation}
r_1^i=\frac{\norm{\tilde{{\bm h}}_1^i}}{\norm{{\bm h}_0^i}}=\norm {\tilde{{\bm h}}_1^i} \,,
\end{equation}
since all ${\bm h}_0^i$ is normalized to $1$ already. An orthonormal basis after time $\Delta t$ is therefore given as  $({\bm h}_1^1\,,{\bm h}_1^2\,, \cdots\,, {\bm h}_1^{2N})$ where
\begin{equation}
{\bm h}_1^i=\frac{1}{r_1^i} \tilde{{\bm h}}_1^i \,.
\end{equation}
This procedure defines the time evolution after one step of $\Delta t$\,. After $n\Delta t$ steps we may write the Lyapunov exponents as
\begin{equation}
\lambda_i=\lim_{n\to\infty}\frac{1}{n\Delta t}\sum\limits_{k=1}^n\log(r_k^i)\,.
\end{equation}
The set $\{\lambda_1,...,\lambda_{2N}\}$ is called the Lyapunov spectrum and as a consequence of this construction $\lambda_1$ is the largest Lyapunov exponent, since ${\bm h}_i^1$ leads toward the direction in the phase space which is most sensitive to the initial conditions and therefore the expansion rate $r_i^1$ has the largest value in this region of the phase space.

A Matlab code solving the Hamilton's equations of motion and evaluating the Lyapunov exponents according to the procedure outlined above is used for our numerical calculations. Let us note that since the phase space is $4$-dimensional there are only four Lyapunov exponents, two of which are zero and the remaining two may be denoted as $\lambda_L$ and $-\lambda_L$ in view of our earlier remarks in this section.

We pick the initial conditions that are used both in the evaluation of the Lyapunov spectrum and the Poincar\'{e} sections as follows. Both $p_r$ and $\theta$ are initially taken to be equal to zero. Using the Hamiltonian \eqref{Hamiltoniannewcoord},  $p_\theta$ can be expressed as
\begin{equation}
p_\theta = \frac{1}{\sqrt{2}}\sqrt{-\frac{1}{2} \kappa^2 r^4-0.1r^3 - \kappa p_\phi r^2 + r^2 -\frac{p_\phi^2}{2}} \,.
\label{in1}
\end{equation}
where we have already set $E =1$ and $\hbar =0.1$. We determine the intervals of $r$ values, which make the argument of the square root in \eqref{in1} positive and restrict to the one in which $r  > 0$. Initial value of  $r$ is chosen randomly from this interval and the initial value of $p_\theta$ is then determined from \eqref{in1}.

We run the Matlab code evaluating the largest positive Lyapunov exponent for $120$ randomly selected initial conditions according to this procedure at each value of $\kappa$ and take their average to obtain the each data point. The error bars are obtained by computing the mean square variances. In the simulation, we take a time step of  $0.5$ and run the code from time $0$ to $3000$. The $\lambda_L$ obtained in this manner is recorded at several values of $p_\phi$ and $\kappa$.

\section{Generic form of the Hamiltonian and the Pure Chern-Simons Limit}\label{Ham1}

Restoring the YM coupling we may write the YMCS action in the form
\be
S_{YMCS} = \frac{1}{g^2} \int {\dd t  \, \Tr \left ( \dfrac{1}{2}(D_0X_i)^2+\dfrac{1}{4}\comm{X_i}{X_j}^2 \right ) } + \kappa \int dt \, \Tr \epsilon_{ij} X_i  \,( D_0 X_j) \,.
\ee
Let us note that the dimensional analysis shows that $g$ has the units $[Length]^{-\frac{3}{2}}$, $\kappa$ has the units $[Length]^2$, while the fields $X_i$ and $A_0$ are of dimension $[Length]^{-1}$. In the paper we have scaled $t \rightarrow g^{-\frac{2}{3}} t $, $\partial_0 \rightarrow g^{\frac{2}{3}} \partial_0 $, $A_0 \rightarrow g^{\frac{2}{3}} A_0 $, $X_i \rightarrow g^{\frac{2}{3}} X_i $ and $\kappa \rightarrow g^{-\frac{4}{3}} \kappa$ to work in the units where $g$ is set to unity.

With the YM coupling included explicitly using (\ref{carLag})  we can write 
\begin{equation}
	S_{YMCS} = \int dt \, \frac{1}{2g^2}(\dot{\va{x}}_1^2 + \dot{\va{x}}_2^2)  + \kappa(\va{x}_1 \vdot \dot{\va{x}}_2 - \va{x}_2 \vdot \dot{\va{x}}_1) - \dfrac{1}{g^2}(\va{x}_1 \cross \va{x}_2)^2 \,,
	\label{carLagG}
\end{equation} 
while the corresponding conjugate momenta and the Hamiltonian are given as 
\be
\va{p}_1 = \dfrac{1}{g^2}\dot{\va{x}}_1 - \kappa\va{x}_2 \,,  \quad \va{p}_2 = \dfrac{1}{g^2}\dot{\va{x}}_2 + \kappa\va{x}_1 \,.
\label{cmomentaG}
\ee
\begin{equation}
	H= \frac{g^2}{2}(\va{p}_1^2 + \va{p}_2^2) + \frac{1}{2}g^2\kappa^2(\va{x}_1^2 + \va{x}_2^2) + g^2\kappa(\va{p}_1\vdot\va{x}_2 - \va{p}_2\vdot\va{x}_1) + \dfrac{1}{g^2}(\va{x}_1 \cross \va{x}_2)^2 \,,
	\label{HamiltonianG}
\end{equation}
We may obtain the pure CS limit by letting $g \rightarrow \infty$. In this limit, we see from (\ref{cmomentaG})  that $\va{p}_1 \rightarrow - \kappa\va{x}_2$ and $\va{p}_2 \rightarrow  \kappa\va{x}_1$. Substituting this in the Hamiltonian (\ref{HamiltonianG}) we find 
\be
H = g^2 (\va{x}_1^2 + \va{x}_2^2) -g^2 (\va{x}_1^2 + \va{x}_2^2)  = 0 \,.
\label{HG}
\ee
Therefore we see that  in the pure CS limit the Hamiltonian vanishes and hence no dynamics or chaos remains. This result is consistent with the general considerations arising from the Chern-Simons theories, as it is known that the pure CS theory has no dynamics, but non-trivial dynamics emerges from coupling to dynamical matter fields, or by considering the CS action on a manifold with boundaries \cite{Dunne}. 

More generally, for a system with generalized coordinates $q_i$ and velocities $\dot{q}_i$, Lagrangian involving first order time derivatives have the generic form
\begin{equation}\label{genL}
L(q_i,\dot{q}_i, t) = \frac{1}{2}g_{ij}\dot{q}_i\dot{q}_j + f_i \dot{q}_i -V \,,
\end{equation}
where $g_{ij}$ is the metric, $f_i$ is some function of the generalized coordinates i.e. $f_i\equiv f_i(q_j)$ and $V$ is a potential $V\equiv V(q_i)$. Canonical momenta are evaluated as
\begin{equation}
p_i=\pdv{L}{\dot{q}_i}=g_{ij}\dot{q}_j+f_i\,.
\end{equation}
In terms of $p_i$, $\dot{q}_i$ can be solved using the inverse metric in the form
\begin{align}
\begin{split}
\dot{q}_i&=g_{ij}^{-1}(p_j - f_j)\,,\\
&=g^{ij}(p_j - f_j)\,.
\end{split}
\end{align}
The Hamiltonian then takes the form
\begin{align}
\begin{split}
H &= p_i \dot{q}_i - L\\
&=p_i g_{ij}^{-1}(p_j - f_j) - \frac{1}{2}g_{ij}g_{ik}^{-1}(p_k - f_k)g_{jl}^{-1}(p_l- f_l) - f_i g_{ij}^{-1}(p_j - f_j) + V\\
&=p_i g_{ij}^{-1}(p_j - f_j) - \frac{1}{2}g_{jl}^{-1}(p_j - f_j)(p_l- f_l) - f_i g_{ij}^{-1}(p_j - f_j) + V\\
&=p_i g_{ij}^{-1}(p_j - f_j) - \frac{1}{2}g_{ij}^{-1}(p_i - f_i)(p_j- f_j) - f_i g_{ij}^{-1}(p_j - f_j) + V\\
&=g_{ij}^{-1}(p_ip_j - p_if_j - \frac{1}{2}p_ip_j + p_if_j - \frac{1}{2}f_if_j - p_if_j + f_if_j) + V\\
&=\frac{1}{2}g_{ij}^{-1}p_ip_j + \frac{1}{2}g_{ij}^{-1}f_if_j - g_{ij}^{-1}p_if_j + V \,,
\end{split}
\label{genham2}
\end{align}
given in \eqref{genericham}.

In order to discuss the limit in which the Lagrangian consists of only first order time derivatives, we may first set $g_{ij} \rightarrow \frac{1}{g^2} g_{ij} $, $g_{ij}^{-1} \rightarrow g^2 g_{ij}^{-1}$ and $V \rightarrow \frac{1}{g^2} V$. We may, therefore, write the generic form of the Hamiltonian as 
\be
H = g^2 ( \frac{1}{2} g_{ij}^{-1} p_i p_j + \frac{1}{2}g_{ij}^{-1}f_if_j  - g_{ij}^{-1}p_if_j ) + \frac{1}{g^2} V \,,
\label{genericHg}
\ee
while the conjugate momenta take the form $p_i = \frac{1}{g^2} g_{ij} \dot{q}_j+f_i$. Thus, in the limit with $g \rightarrow \infty$ we have, $p_i \rightarrow f_i$ and upon substituting this in (\ref{genericHg}) we immediately see that $H=0$. 

It is somewhat more subtle to see the pure CS limit from the Hamiltonian given in the angular coordinates in (\ref{Hamiltoniannewcoord}). We will provide a concrete discussion of this limit in the next appendix.

\section{Derivation of the Hamiltonian in the New Coordinates}

\subsection{Metric}\label{asec:metric}

A general $SO(3)$ element in the  Euler's parametrization with $z-x-z$ active rotation with the angles $\alpha, \beta, \gamma$ respectively is given by \cite{Goldstein}
\begin{align}
	\begin{split}
		R(\alpha,\beta,\gamma) =
		\begin{pmatrix}
			c(\alpha)c(\gamma) - s(\alpha)c(\beta)s(\gamma) &
			-s(\alpha)c(\beta)c(\gamma) - c(\alpha)s 
			(\gamma) &s(\alpha)s(\beta) \\
			c(\gamma)s(\alpha) + c(\alpha)c(\beta)s(\gamma) &
			c(\alpha)c(\beta)c(\gamma) - s(\alpha)s(\gamma) & 
			- c(\alpha)s(\beta) \\
			s(\beta)s(\gamma) &s(\beta)c(\gamma) &c(\beta) 
		\end{pmatrix},
	\end{split}
\end{align}
where $s$ and $c$ stand for sine and cosine, respectively. This can be facilitated to obtain the matrix $M$ in \ref{newcoord}. 

The metric in the new coordinates $(r,\theta,\phi,\alpha,\beta,\gamma)$  is evaluated using $g_{ij}=\Tr(\partial_i M^\dagger \partial_j M)$ and it yields
\begin{equation}
g_{ij}=
\begin{pmatrix}
1 & 0 & 0 & 0 & 0 & 0 \\
0 & \frac{r^2}{2} & \frac{1}{2} r^2 \sin (\theta ) & \frac{1}{2} r^2 \cos (\beta ) & 0 & \frac{r^2}{2} \\
0 & \frac{1}{2} r^2 \sin (\theta ) & r^2 & r^2 \cos (\beta ) \sin (\theta ) & 0 & r^2 \sin (\theta ) \\
0 & \frac{1}{2} r^2 \cos (\beta ) & g_{34} &g_{44} & g_{45} & r^2 \cos (\beta ) \\
0 & 0 & 0 & g_{54} & g_{55} & 0 \\
0 & \frac{r^2}{2} & r^2 \sin (\theta ) & r^2 \cos (\beta ) & 0 & r^2 \\
\end{pmatrix}\,,
\end{equation}
where
\begin{align}
\begin{split}
g_{34}=&r^2 \cos (\beta ) \sin (\theta )\,,\\
g_{44}=&-\frac{1}{8} r^2 \cos (2 \beta ) \cos (2 (\gamma +\theta ))-\frac{1}{16} r^2 \cos (2 (\beta -\gamma ))\\
&-\frac{1}{16} r^2 \cos (2 (\beta +\gamma ))+\frac{1}{4} r^2 \cos (2 \beta )\\
&+\frac{1}{4} r^2 \cos (\theta ) \cos (2 \gamma +\theta )+\frac{3 r^2}{4}\,,\\
g_{54}=&-\frac{1}{2} r^2 \sin (\beta ) \cos (\theta ) \sin (2 \gamma +\theta )\,,\\
g_{45}=&-\frac{1}{2} r^2 \sin (\beta ) \cos (\theta ) \sin (2 \gamma +\theta )\,,\\
g_{55}=&-\frac{1}{4} r^2 \cos (2 (\gamma +\theta ))-\frac{1}{4} r^2 \cos (2 \gamma )+\frac{r^2}{2}\,.
\end{split}
\end{align}
The inverse metric $g^{-1}_{ij}$ is given as 
\begin{equation}
g^{ij} = 
\begin{pmatrix}
1 & 0 & 0 & 0 & 0 & 0 \\
0 & \frac{4}{r^2} & 0 & 0 & 0 & -\frac{2}{r^2} \\
0 & 0 & \frac{\sec ^2(\theta )}{r^2} & 0 & 0 & -\frac{\sec (\theta ) \tan (\theta )}{r^2} \\
0 & 0 & 0 & g^{44} & g^{45} & g^{46} \\
0 & 0 & 0 & g^{54} & g^{55} & g^{56} \\
0 & -\frac{2}{r^2} & -\frac{\sec (\theta ) \tan (\theta )}{r^2} & g^{64} & g^{65} & g^{66}\\
\end{pmatrix}\,,
\label{invsmtrc}
\end{equation}
where
\begin{align}
\begin{split}
g^{44}&=-\frac{(\cos (2 \gamma )+\cos (2 (\gamma +\theta ))-2) \csc ^2(\beta ) \csc ^2(\theta )}{r^2}\,,\\
g^{45}&=\frac{2 \cot (\theta ) \csc (\beta ) \csc (\theta ) \sin (2 \gamma +\theta )}{r^2}\,,\\
g^{46}&=\frac{(\cos (2 \gamma )+\cos (2 (\gamma +\theta ))-2) \cot (\beta ) \csc (\beta ) \csc ^2(\theta )}{r^2}\,,\\
g^{54}&=\frac{2 \cot (\theta ) \csc (\beta ) \csc (\theta ) \sin (2 \gamma +\theta )}{r^2}\,,\\
g^{55}&=\frac{(\cos (2 \gamma )+\cos (2 (\gamma +\theta ))+2) \csc ^2(\theta )}{r^2}\,,\\
g^{56}&=-\frac{2 \cot (\beta ) \cot (\theta ) \csc (\theta ) \sin (2 \gamma +\theta )}{r^2}\,,\\
g^{64}&=\frac{(\cos (2 \gamma )+\cos (2 (\gamma +\theta ))-2) \cot (\beta ) \csc (\beta ) \csc ^2(\theta )}{r^2}\,,\\
g^{65}&=-\frac{2 \cot (\beta ) \cot (\theta ) \csc (\theta ) \sin (2 \gamma +\theta )}{r^2}\,,\\
g^{66}&=\frac{-(\cos (2 \gamma )+\cos (2 (\gamma +\theta ))-2) \cot ^2(\beta ) \csc ^2(\theta )+\sec ^2(\theta )+1}{r^2}\,.\\
\end{split}
\end{align}

\subsection{Hamiltonian in the new coordinates}

Corresponding to the generalized coordinates $(r, \theta, \phi, \alpha, \beta, \gamma)$, we label the associated conjugate momenta as $(p_r, p_\theta, p_\phi, p_\alpha, p_\beta, p_\gamma)$. Using the inverse metric in \ref{invsmtrc}, we have the first term in the generic form of the Hamiltonian \eqref{genericham} (or\eqref{genham2}) given as
\begin{align}
	\begin{split}
		\frac{1}{2}g_{ij}^{-1}p_ip_j =&
		-\frac{\csc ^2(\beta ) \cos (2 \gamma ) \csc ^2(\theta ) p_{\alpha }^2}{2 r^2}\\
		&-\frac{\csc ^2(\beta ) \csc ^2(\theta ) p_{\alpha }^2 \cos (2 (\gamma +\theta ))}{2 r^2}\\
		&-\frac{2 \cot (\beta ) \csc (\beta ) \csc ^2(\theta ) p_{\alpha } p_{\gamma }}{r^2}\\
		&+\frac{\cot (\beta ) \csc (\beta ) \cos (2 \gamma ) \csc ^2(\theta ) p_{\alpha } p_{\gamma }}{r^2}\\
		&+\frac{\cot (\beta ) \csc (\beta ) \csc ^2(\theta ) p_{\alpha } p_{\gamma } \cos (2 (\gamma +\theta ))}{r^2}\\
		&+\frac{2 \csc (\beta ) \cot (\theta ) \csc (\theta ) p_{\alpha } p_{\beta } \sin (2 \gamma +\theta )}{r^2}\\
		&+\frac{\csc ^2(\beta ) \csc ^2(\theta ) p_{\alpha }^2}{r^2}+\frac{\cot ^2(\beta ) \csc ^2(\theta ) p_{\gamma }^2}{r^2}\\
		&-\frac{\cot ^2(\beta ) \cot (\theta ) \csc (\theta ) p_{\gamma }^2 \cos (2 \gamma +\theta )}{r^2}\\
		&+\frac{\cot (\theta ) \csc (\theta ) p_{\beta }^2 \cos (2 \gamma +\theta )}{r^2}\\
		&-\frac{2 \cot (\beta ) \cot (\theta ) \csc (\theta ) p_{\beta } p_{\gamma } \sin (2 \gamma +\theta )}{r^2}\\
		&+\frac{\csc ^2(\theta ) p_{\beta }^2}{r^2}-\frac{\tan (\theta ) \sec (\theta ) p_{\gamma } p_{\phi }}{r^2}\\
		&-\frac{2 p_{\gamma } p_{\theta }}{r^2}+\frac{\sec ^2(\theta ) p_{\gamma }^2}{2 r^2}+\frac{p_{\gamma }^2}{2 r^2}\\
		&+\frac{\sec ^2(\theta ) p_{\phi }^2}{2 r^2}+\frac{2 p_{\theta }^2}{r^2}+\frac{p_r^2}{2}\,.
	\end{split}
\label{HT1}
\end{align}
In order to proceed, we need to evaluate the form of $\va{f}_i = - \kappa \varepsilon_{ij} \va{x}_j$ ($i,j:1,2$)  in the new coordinates. The function $f_i=f_i(q_j)$ and $\dot{q}_i$ in the Lagrangian \eqref{carLag} appear as
\begin{equation}\label{45}
f_i \dot{q}_i \equiv \kappa \va{x}_1\vdot\dot{\va{x}}_2 - \kappa\va{x}_2\vdot\dot{\va{x}}_1\,.
\end{equation}
Since the $i^{th}$ column of the matrix $M$ in \ref{newcoord} correspond to the components of $\va{x}_i$, and so does the correspondence goes with their time derivatives, right hand side of \eqref{45} can be written by taking the inner products of the column vectors of $M$ and $\dot{M}$ and this yields
\begin{equation}\label{hao}
	f_i \dot{q}_i = - \frac{1}{2}r^2\kappa\left(2\dot{\phi}+\sin(\theta)(2\dot{\alpha}\cos(\beta)+2\dot{\gamma}+\dot{\theta})\right)\,.
\end{equation}
Since $f_i \dot{q}_i = f_1\dot{r} + f_2\dot{\theta} + f_3 \dot{\phi} + f_4 \dot{\alpha} + f_5\dot{\beta} + f_6\dot{\gamma}$ in the new coordinates, the coefficients  $f_i(r,\theta,\phi,\alpha,\beta,\gamma)$, $(i:1,...,6)$ are now easily read out from \eqref{hao} to be 
\begin{align}
	\begin{split}
		f_1&=0\,,\\
		f_2&=-\frac{1}{2}r^2\kappa\sin(\theta)\,,\\
		f_3&=-r^2\kappa\,,\\
		f_4&=-r^2\kappa\sin(\theta)\cos(\beta)\,,\\
		f_5&=0\,,\\
		f_6&=-r^2\kappa\sin(\theta)\,.
	\end{split}
\label{fs}
\end{align}

With $f_i$ given \eqref{fs}, we can evaluate  the second and the third term in \eqref{genericham}. We find
\begin{equation}
\frac{1}{2} g_{ij}^{-1} f_i f_j = \frac{1}{2} r^2 \kappa^2\,,
\label{HT2}
\end{equation}
and
\begin{equation}
\frac{1}{2} g_{ij}^{-1} p_i f_j = -p_\phi \kappa\,,
\label{HT3}
\end{equation}
The last term takes the form
\begin{equation}
\frac{1}{2}(\va{x}_1 \cross \va{x}_2)^2 = \frac{1}{4}r^4\sin[2](\theta)\,,
\label{HT4}
\end{equation}
which is the same as that would be obtained had we used the matrix $M^0$, since the square of the cross product of the column vectors of $M$ is a scalar and does not get affected by gauge rotations. 

Putting \eqref{HT1}, \eqref{HT2}, \eqref{HT3}, \eqref{HT4} together and imposing the Gauss law constraint $	\va{L}=0 $ via \eqref{gausslaw4} as\footnote{This is proved in the next appendix.}
\be
p_\alpha = p_\beta = p_\gamma =0 \,.
\ee
we finally obtain the Hamiltonian given in  \eqref{Hamiltoniannewcoord}.

\subsection{Pure CS limit in the new coordinates}

With the YM coupling written explicitly, the Hamiltonian in the angular coordinates is given as
\be
H = g^2\left(\frac{1}{2}p_r^2 + \frac{2}{r^2}p_\theta^2 + \frac{p_\phi^2}{2r^2\cos[2](\theta)} + \kappa p_\phi  + \frac{\kappa^2 r^2}{2}\right) + \frac{1}{4g^2}r^4\sin[2](\theta) \,. 
\label{HamiltoniannewcoordG}
\ee
Since the Gauss Law constraint (\ref{gausslaw4}) $p_\alpha = p_\beta=p_\gamma = 0$ is already imposed in (\ref{HamiltoniannewcoordG}) it is somewhat more subtle to see the pure CS limit. The formal way to proceed is to use the form of the Hamiltonian prior to imposing the Gauss law constraint which is given by (\ref{genericHg}) with (\ref{HT1}),(\ref{HT2}) and (\ref{HT3}). As $g \rightarrow \infty$ we have $p_i \rightarrow f_i$ with $f_i$ given in (\ref{fs}), and a short calculation in Mathematica confirms that $H=0$. A rather quick way to see this result from (\ref{Hamiltoniannewcoord}) is as follows. From the Gauss law constraint the equations $p_\alpha=0$, $p_\gamma =0$ imply that $\theta=0$. This gives, $p_\theta = f_2 = - \frac{1}{2} \kappa r^2 \sin \theta =0$. We further have $p_r = f_1$ and $f_1=0$ and, $p_\phi = f_3$, $f_3 = -r ^2 \kappa$. Substituting these in (\ref{Hamiltoniannewcoord}) we immediately obtain $H=0$.

\section{Angular Momentum Vector in terms of Euler Angles and Conjugate Momenta}\label{angularmomentum}

Angular velocities can be expressed in terms of Euler angles and their time derivatives as \cite{Goldstein, Marsden}
\begin{align}\label{weq}
	\begin{split}
		w_1 &= \dot{\gamma } \sin (\alpha ) \sin (\beta )+\dot{\beta } \cos (\alpha )\,,\\
		w_2 &= \dot{\beta } \sin (\alpha )-\dot{\gamma } \cos (\alpha ) \sin (\beta )\,,\\
		w_3 &= \dot{\alpha }+\dot{\gamma } \cos (\beta )\,.
	\end{split}
\end{align}

In terms of Euler angles and their time derivatives rotational kinetic energy takes the form
\begin{align}
	\begin{split}
		T=&\frac{1}{2}(I_1 w_1^2 + I_2 w_2^2 + I_3 w_3^2)\\
		=&I_1(\dot{\gamma }^2 \sin ^2(\alpha ) \sin ^2(\beta )+2 \dot{\beta } \dot{\gamma } \sin (\alpha ) \cos (\alpha ) \sin (\beta )+\dot{\beta }^2 \cos ^2(\alpha ))\\
		&+I_2(\dot{\gamma }^2 \cos ^2(\alpha ) \sin ^2(\beta )-2 \dot{\beta } \dot{\gamma } \sin (\alpha ) \cos (\alpha ) \sin (\beta )+\dot{\beta }^2 \sin ^2(\alpha ))\\
		&+I_3(2 \dot{\alpha } \dot{\gamma } \cos (\beta )+\dot{\alpha }^2+\dot{\gamma }^2 \cos ^2(\beta ))\,,
	\end{split}
\end{align}
where $I_i$ are the moment of inertia with respect to the principal axes associated to $z-x-z$ active rotation with the angles $\alpha, \beta, \gamma$.

Momentum conjugate to the Euler angles $\alpha, \beta, \gamma$ are
\be
p_\alpha=\pdv{T}{\dot{\alpha}}\,,\,p_\beta=\pdv{T}{\dot{\beta}}\,,\,p_\gamma=\pdv{T}{\dot{\gamma}} \,,
\ee
and we have 
\begin{equation}\label{note}
	\begin{pmatrix}
		p_\alpha\\
		p_\beta\\
		p_\gamma
	\end{pmatrix}
	=
	\begin{pmatrix}
		0 & I_3 & I_3 \cos (\beta ) \\
		0 & A_{22} & A_{23} \\
		I_3 \cos (\beta ) & A_{32} & A_{33}
	\end{pmatrix}
	\begin{pmatrix}
		\dot{\alpha}\\
		\dot{\beta}\\
		\dot{\gamma}
	\end{pmatrix}
\end{equation}
where the the remaining components of the matrix $A$ are given as
\begin{align}
	\begin{split}
		A_{22}&=I_1 \cos ^2(\alpha )+I_2 \sin ^2(\alpha )\,,\\
		A_{23}&=I_1 \cos (\alpha ) \sin (\alpha ) \sin (\beta )-I_2 \cos (\alpha ) \sin (\alpha ) \sin (\beta )\,,\\
		A_{32}&=I_1 \cos (\alpha ) \sin (\alpha ) \sin (\beta )-I_2 \cos (\alpha ) \sin (\alpha ) \sin (\beta )\,,\\
		A_{33}&=I_3 \cos ^2(\beta )+I_2 \cos ^2(\alpha ) \sin ^2(\beta )+I_1 \sin ^2(\alpha ) \sin ^2(\beta ) \,.
	\end{split}
\end{align}
Writing \eqref{note} as $P = A \, {\Theta}$ in short, the column matrix $\dot{\Theta} = (\dot{\alpha},\dot{\beta},\dot{\gamma})^T $ can be obtained from the equation $\dot{\Theta}=A^{-1} \, P$\,. Substituting $\dot{\alpha},\dot{\beta},\dot{\gamma}$ obtained from this equation into \eqref{weq} yields $w_1,w_2,w_3$ in terms of $p_\alpha,p_\beta,p_\gamma$. Finally, we obtain the components of angular momentum $\va{L}$ using $L_i=\pdv{T}{w_i}=I_iw_i$ (no sum over $i$) as \cite{Marsden}
\begin{equation}
	\va{L}=
	\begin{pmatrix}
		\sin (\alpha ) (p_\gamma \csc (\beta )-p_\alpha \cot (\beta ))+p_\beta \cos (\alpha )\\
		\cos (\alpha ) \csc (\beta ) (p_\alpha \cos (\beta )-p_\gamma)+p_\beta \sin (\alpha )\\
		p_\alpha
	\end{pmatrix}\,.
\end{equation}

Therefore, the Gauss Law constraint $\va{L} = 0$ is equivalent to \eqref{gausslaw4}.

\end{document}